\documentclass{aastex63}

\usepackage{amsmath}
\usepackage{enumitem}
\usepackage{graphics,graphicx}
\usepackage{epsfig}

\usepackage{subfigure}
\usepackage{longtable}
\usepackage{times}

\begin{document}


\title{Proposing a Physical Mechanism to Explain Various Observed Sources of QPOs
  by Simulating the Dynamics of Accretion Disks around the Black Holes} 

\author{Orhan Donmez}
\altaffiliation{College of Engineering and Technology, American
  University of the Middle East, Egaila 54200, Kuwait}

\begin{abstract}
We propose a mechanism to explain the low-frequency $QPOs$ observed in $X$-ray binary systems and AGNs. To achieve this, we perturbed stable accretion disks around Kerr and EGB black holes at different angular velocities, revealing characteristics of shock waves and oscillations on the disk. By applying this perturbation to scenarios with varying alpha values for EGB black holes and different spin parameters for Kerr black holes, we numerically observed changes in the disk dynamic structure and its oscillations. Through various numerical models, we found that the formation of one- and two-armed spiral shock waves on the disk serves as a mechanism for generating $QPOs$. We compared the $QPOs$ obtained from numerical calculations with the low-frequency $QPOs$ observed in $X$-ray binary systems and AGN sources, finding high consistency with observations. We observed that the shock mechanism, leading to $QPOs$, explains the $X$-ray binaries and AGNs studied in this article. Our numerical findings indicate that $QPOs$ are more strongly dependent on the EGB constant than on the black hole spin parameter. However, we highlighted that the primary impact on oscillations and $QPOs$ is driven by the perturbation angular velocity. The results from the models showed that the perturbation asymptotic speed at $V_{\infty}=0.2$ independently generates $QPO$ frequencies, regardless of the black hole spin parameter and the EGB coupling constant. Therefore, for the moderate value of $V_{\infty}$, a two-armed spiral shock wave formed around the black hole is suggested as a decisive mechanism in explaining low-frequency $QPOs$.
\end{abstract}



\section{Introduction}
\label{Introduction}

Quasi-Periodic Oscillations ($QPOs$) studies for the black holes and neutron stars have been extensively examined in the literature through theoretical \citep{Kato2002PASJ, Abramowicz2001A&A, Abramowicz2013, Ingram2019}, numerical \citep{Schnittman2006ApJ, Luciano2015GReGr, CruzOsorio2023JCAP, Donmez2023arXiv231013847D}, and observational approaches \citep{Smith2021ApJ}. The presence of a black hole at the center of the observed sources confirms the existence of strong gravitational attraction, leading to high-frequency $QPOs$ resulting from the intense gravitational forces near the black hole horizon, affecting the disk structure or the precession motion of the generated jet. The origin of $QPOs$ may stem from various factors such as the Keplerian motion of matter on the disk, the rotation parameter of the black hole ($a/M$), the influence of alternative gravitational theories (e.g., the coupling constant of EGB), and general relativistic effects like the Lense-Thirring effect, or a combination of these effects. Such factors are known to create shock waves or cavities on the disk near the black hole, leading to oscillations as matter continuously falls toward the black hole \citep{vanderKlis2000ARA&A, Ingram2019, CruzOsorio2020ApJ, Donmez2023arXiv231013847D}.

Based on theoretical and numerical results, the physical mechanisms leading to the formation of $QPOs$ observed in Active Galactic Nuclei ($AGNs$) have been widely discussed in the literature \citep{King2013MNRAS, Ackermann2015ApJ, Smith2023ApJ}. The $X$-ray source used in the detection of $QPOs$ exhibits significant variations from one source to another. Several reasons could account for this variability. One is the non-ballistic motion of the emitted matter \citep{Li2016PASP}. Another reason could be the response of a low-frequency $QPO$ source with a jet to the instabilities within the jet \citep{Ferreira2022A&A}. In this model, it is believed that disk instabilities lead to the wobbling of the jet. As a result of this wobbling, emission occurs in both directions of the jet. Depending on the nature of the jet wobble, there could be interactions between the jet and the disk. Such interactions might cause instabilities on the disk, leading to the modulation of the Kepler frequency or the Lense-Thirring precession frequency. Combined with the rotation of the  black hole, this effect would cause the disk to undergo $QPOs$, altering the disk pressure and potentially generating instabilities \citep{Tchekhovskoy2011MNRAS}, such as Kelvin-Helmholtz instability due to the pressure force \citep{Porth2015MNRAS}. While this model was initially proposed for $X$-ray binary systems, it is also applicable to $AGNs$, where these types of instabilities can modulate the jets.

The influence of high-frequency oscillations on the mass and spin parameter of black holes has been investigated. The study indicates that massive black holes behave differently compared to stellar-mass black holes. It has been found that high-frequency $QPOs$ from massive black holes cannot be explained by known mechanisms. Conversely, the $QPOs$ observed around stellar-mass black holes can be accounted for. Therefore, theoretical models considered as physical mechanisms for high-frequency $QPOs$, such as orbital resonance, warped disk, and disk-jet interaction, are capable of generating $QPOs$ around stellar-mass black holes but are found to be incompatible with massive black holes \citep{Smith2021ApJ}.

Applying modified gravity (MOG) to black hole-disk interaction problems contributes to explaining observations \citep{Kolo2020EPJC}. In this context, by defining the black hole with MOG, researchers have investigated how the inner radius of the disk around the black hole, the last stable orbit, and the $QPOs$ change with the MOG parameter. They then examined the radial oscillation frequencies due to changes in the radial direction of the last stable rotation of the disk. According to the results obtained, it was found that the resonance radius generated by MOG black holes is larger than that of Kerr, as expected. As the MOG parameter increases, this radius moves away from the black hole. They calculated geodesic resonance frequencies and used them to compare with the microquasars $GRO$ $1655-40$, $XTE$ $1550-564$, and $GRS$ $1915+105$, finding that some models aligned with the observational results.

Observations of $X$-ray binaries and $AGNs$ with black holes at their centers have revealed a wide range of $QPO$ oscillation frequencies. Various known physical factors and parameters contribute to these broad-spectrum frequencies \citep{Thomson2014PhDT, Dhang2018MNRAS}. One such factor is the oscillations near the surface of the black hole, especially due to the warping of space-time caused by the black hole rotation parameter \citep{Kolo2020EPJC}. Additionally, the different physical parameters associated with alternative gravities defining the space-time within the black hole also play a role \citep{Donmez2023arXiv231013847D}. In this context, we will discuss the impact of $a/M$ on $QPO$ frequencies, although not from an extremely broad perspective. Furthermore, we will examine the effect of the EGB coupling constant ($\alpha$), the most influential parameter aside from $a/M$, in the context of EGB gravity on $QPO$ frequencies. Most importantly, we will explore how matter falling towards the black hole, which contains a stable accretion disk in its initial state, contributes to the formation of shock waves and $QPOs$. As will be evident in the subsequent sections of the article, it has been observed that matter falling towards the black hole interacts with the stable disk, resulting in the formation of one- and two-armed spiral shock waves. The structure of these waves and the physical parameters of perturbation have been found to significantly impact the stability of the disk and, consequently, the observed $QPO$ frequencies. Therefore, this paper will define physical mechanisms in accordance with the disk structure obtained through numerical calculations. The compatibility or divergence of the frequencies obtained through observations with numerical results will be discussed, and their physical reasons will be explored.

Here, we propose a mechanism that can generate the low-frequency oscillations observed in $X$-ray binaries and AGNs. Determined solely through numerical modeling, this mechanism involves both one- and two-armed shock waves, which significantly impact the disk stability. Changes in the angular velocity of the perturbation, which cause the formation of these shock waves, have a profound effect on both $QPOs$ and disk stability. In the following sections, we will discuss how angular velocity affects the dynamic structure of the disk, the conditions for the formation of instabilities and their transition to $QPOs$, the compatibility of low-frequency $QPOs$, among other details.

In this paper, we attempt to explain the low-frequency $QPOs$ observed from various sources through the physical mechanism arising from perturbations of the stable disk around black holes. To propose a mechanism aligned with the observational results, we carry out the following steps: In section \ref{GRHE1}, we provide a summary of the Kerr and EGB gravitational metrics, which describe the properties of black holes and the geometry of the surrounding space, citing relevant references. We also summarize the General Relativistic Hydrodynamics (GRH) equations we solve to obtain numerical results. In section \ref{InitialBC}, we present the formation of stable disks used in all models and the boundary values, supported with references. We illustrate the variations in disk structures and resulting shock waves due to modeling stable disks around black holes with different perturbation parameters. This analysis accounts for the black hole spin and the EGB coupling constant in section \ref{dynamics}. In section \ref{instability}, we discuss how the $m=1$ and $m=2$ modes of instability in the disk vary between different models and explore the characterization of the instability using root mean square (RMS). In section \ref{QPO}, we present the frequencies of $QPOs$ on the disk by conducting power spectrum analysis, showing the frequencies for different models. In section \ref{QPOs_source}, we compare the frequencies obtained in section \ref{QPO} with observed $QPO$ frequencies from various sources, leading to the proposal of potential physical mechanisms. These comparisons are made for selected sources from both $X$-ray binaries and AGNs. In section \ref{Conclusion}, we summarize the alignment of numerically obtained frequencies with observed frequencies, the dependence on physical parameters used in numerical models, and suggest physical mechanisms that can explain the observed frequencies.


\section{Governing Equations and Metrics}
\label{GRHE1}

We solve the GRH equations numerically to unveil the properties of black hole-disk interactions, as well as to study instabilities, various types of shock waves, and the physical properties of radiation emitted from the disk. Through solving these equations, we can describe the stable accretion disks that form around black holes and model any new scenarios that may arise from perturbations.

The GRH equations are derived from the principles of general relativity and classical hydrodynamics. They incorporate the static spacetime metric solution obtained from the Einstein equations. Initially derived in a non-conservation form, the GRH equations are then transformed into a conservation form using the space-time foliation and the stress-energy tensor ($T^{ab}$) of matter, which is assumed to exhibit ideal behavior. Solving the GRH equations with high-resolution numerical methods reveals the behavior of matter near the  black hole under strong gravitational forces. Consequently, they are expressed in conservation form. The conservation form of the 2D GRH equations is as follows \citep{Donmez1, Donmez2, Donmez5}:

\begin{eqnarray}
  \frac{\partial U}{\partial t} + \frac{\partial F^r}{\partial r} +
  \frac{\partial F^{\phi}}{\partial \phi}
  = S,
\label{GRHEq1}
\end{eqnarray}

\noindent
where $U$ represents the conserved parameter, $F^r$ and $F^\phi$ represent the fluxes in the $r$ and $\phi$ directions, respectively, and $S$ defines the source term. The stress-energy-momentum tensor of matter exhibiting ideal behavior is $T^{ab} = \rho h u^{a}u^{b} + P g^{ab}$. Here, $\rho$ stands for the rest-mass density, $p$ represents fluid pressure, $h$ indicates specific enthalpy, $u^{a}$ denotes the 4-velocity of the fluid, and $g^{ab}$ serves as the metric for the curved space-time. The indices $a$, $b$, and $c$ take values ranging from $0$ to $3$.

$U$, among the variables in the conservation form of the GRH equations, is given as,

\begin{eqnarray}
  U =
  \begin{pmatrix}
    D \\
    S_j \\
    \tau
  \end{pmatrix}
  =
  \begin{pmatrix}
    \sqrt{\gamma}W\rho \\
    \sqrt{\gamma}h\rho W^2 v_j\\
    \sqrt{\gamma}(h\rho W^2 - P - W \rho)
    \end{pmatrix},
\label{GRHEq2}
\end{eqnarray}

\noindent
where $W = (1 - \gamma_{a,b}v^i v^j)^{1/2}$ is Lorentz factor,
$h = 1 + \epsilon + P/\rho$ is enthalpy, and $\epsilon$ stands for the
internal energy. The three-velocity of the fluid is given by $v^i = u^i/W + \beta^i$.
It is  assumed that when the matter around the black hole interacts with the black hole and
by itself, we assumed an ideal behavior for matter. Therefore, the ideal pressure
equation $P = (\Gamma - 1)\rho\epsilon$ is used in numerical simulations.

On the other hand, the flux and source terms seen in Eq.\ref{GRHEq1} are defined as follows:

\begin{eqnarray}
  \vec{F}^i =
  \begin{pmatrix}
    \tilde{\alpha}\left(v^i - \frac{1}{\tilde{\alpha}\beta^i}\right)D \\
    \tilde{\alpha}\left(\left(v^i - \frac{1}{\tilde{\alpha}\beta^i}\right)S_j + \sqrt{\gamma}P\delta^i_j\right)\\
    \tilde{\alpha}\left(\left(v^i - \frac{1}{\tilde{\alpha}\beta^i}\right)\tau  + \sqrt{\gamma}P v^i\right)
    \end{pmatrix},
\label{GRHEq3}
\end{eqnarray}

\noindent and,

\begin{eqnarray}
  \vec{S} =
  \begin{pmatrix}
    0 \\
    \tilde{\alpha}\sqrt{\gamma}T^{ab}g_{bc}\Gamma^c_{aj} \\
    \tilde{\alpha}\sqrt{\gamma}\left(T^{a0}\partial_{a}\tilde{\alpha} - \tilde{\alpha}T^{ab}\Gamma^0_{ab}\right)
   \end{pmatrix},
\label{GRHEq4}
\end{eqnarray}

\noindent where $\Gamma^c_{ab}$ is the Christoffel symbol.

The matrix $g^{ab}$, appearing in the GRH equations (Eq.\ref{GRHEq1}), is the extended metric tensor that defines the geometry around the black hole. This matrix characterizes not only the physical properties of the black hole but also the curvature of the surrounding space. In other words, by understanding this matrix, we can calculate the strong gravitational field that matter is subjected to during its interaction with the black hole, the trajectory it follows as it falls into the black hole, and various other phenomena.

Using different matrices in numerical calculations provides the opportunity to explore the implications of alternative gravities \citep{Ghosh2, Donmez3, Donmez_EGB_Rot, Donmez2023arXiv231013847D}. Therefore, by solving the same problems within different gravitational theories, we can offer explanations for observational phenomena. In this context, this article employs two different space-time matrices: the Kerr and Einstein-Gauss-Bonnet (EGB) gravities.

The Kerr black hole generates a very strong gravitational force due to its rotation parameter and significantly warps the surrounding space-time. Particularly when studying the behavior of matter in the region known as the inner disk, the Kerr metric should be used. The Kerr black hole in Boyer-Lindquist coordinates is described by the following metric \citep{Donmez6}:

\begin{eqnarray}
  ds^2 = -\left(1-\frac{2Mr}{\sum^2}\right)dt^2 - \frac{4Mra}{\sum^2}sin^2\theta dt d\phi
  + \frac{\sum^2}{\Delta_1}dr^2 + \sum^2 d\theta^2 + \frac{A}{\sum^2}sin^2\theta d\phi^2,
\label{GRHEq5}
\end{eqnarray}

\noindent where $\Delta_1 = r^2 - 2Mr +a^2$, and
$A = (r^2 + a^2)^2 - a^2\Delta sin^2\theta$.
In Boyer-Lindquist coordinates, the shift vector and lapse function of the Kerr metric are
$\beta^i = (0,0,-2Mar/A)$ and $\tilde{\alpha} = (\sum^2 \Delta_1/A)^{1/2} $.

EGB gravity is not a solution to the Einstein field equations but rather a modification or an extension of general relativity. It is derived by incorporating a new term, the Gauss-Bonnet term, into the theory of gravity, thus presenting itself as an alternative gravity theory. Unlike the Kerr solution, this alternative gravity theory is characterized by the alpha ($\alpha$) parameter. The impact of this parameter is particularly significant when examined in regions near black holes and can be utilized to explain observational results. $\alpha$ introduces higher curvature corrections, offering a new perspective on the behavior of black holes in extreme conditions and higher-dimensional space-time \citep{Fernandes2022}. The 4D EGB rotating black hole metric is detailed in \citep{Donmez_EGB_Rot, Donmezetal2022}.

\begin{eqnarray}
  ds^2 &=& -\frac{\Delta_2 - a^2sin^2\theta}{\Sigma}dt^2 + \frac{\Sigma}{\Delta_2}dr^2 -
  2asin^2\theta\left(1- \frac{\Delta_2 - a^2sin^2\theta}{\Sigma}\right)dtd\phi + 
  \Sigma d\theta^2 + \nonumber \\
  && sin^2\theta\left[\Sigma +  a^2sin^2\theta \left(2- \frac{\Delta_2 -  
   a^2sin^2\theta}{\Sigma} \right)  \right]d\phi^2,
\label{GRHEq6}
\end{eqnarray}

\noindent
where $\Sigma$ and $\Delta_2$ are given as $\Sigma = r^2 + a^2 \cos^2\theta$
and $\Delta_2 = r^2 + a^2 + \frac{r^4}{2\alpha}\left(1 - \sqrt{1 +
  \frac{8 \alpha M}{r^3}} \right)$, respectively. Here, $a$, $\alpha$, and $M$
represent the black hole spin parameter, Gauss-Bonnet coupling constant, and mass of the
black hole, respectively. The black hole horizon is determined by solving
the equations $\Delta_1=0$ and $\Delta_2=0$.
The lapse function $\tilde{\alpha}$ and the shift vectors of the $4D$ EGB metric
are given as $\tilde{\alpha} = \sqrt{\frac{a^2(1-f(r))^2}{r^2+a^2(2-f(r))} + f(r)}$
and $\beta^i = \left(0,0, \frac{a r^2}{2\pi \alpha}\left(1 - \sqrt{1 + \frac{8 \pi
    \alpha M}{r^3}}\right)\right)$, respectively. Here, $f(r) = 1 +
\frac{r^2}{2\alpha}\left(1 - \sqrt{1 + \frac{8 \alpha M}{r^3}} \right)$.

The gamma matrix, $\gamma_{i,j}$,  appearing in the GRH equations defines three-dimensional space.
It is determined from $g_{ab}$ for Kerr and EGB gravities.
Latin indices $i$ and $j$ range from $1$ to $3$.

\section{Initial, Boundary Conditions, and Assumptions}
\label{InitialBC}

The structure of the accretion disk around black holes, formed as a result of its interaction with the black hole and perturbations, can be investigated using shock waves and instabilities to explain the observed $QPO$ frequencies. This enables the prediction of physical mechanisms that trigger the observed $QPOs$.

To achieve this, it is crucial for the initial accretion disk around the black hole to be stable. Elucidating the mechanisms resulting from perturbations in the disk necessitates a stable initial state of the disk. As previously done in \citet{Donmez2023, Donmez2023arXiv231013847D}, a stable disk was established by accreting matter spherically towards the black hole. The assumption here is that matter accumulates spherically towards the black hole. Systems harboring black holes at their centers, often formed due to various events like supernova explosions, can attract matter (debris) towards themselves, thus forming a stable disk. The formation of this stable initial disk and the parameters used are extensively detailed in \citet{Donmez2023, Donmez2023arXiv231013847D}.

The stable initial disk used in this article was derived through numerical modeling. This approach is adopted because, as mentioned earlier, two different gravitational theories are employed to understand the physical mechanisms behind the $QPOs$. Neither of these gravitational theories possesses an analytically known stable accretion disk. Therefore, the space-time metrics, which are solutions of Einstein equations or extensions of general relativity, depend on the black hole spin and the EGB coupling constant. Investigating $QPOs$ with respect to these parameters and uncovering the physical mechanisms that can lead to these $QPOs$ is crucial.

In the second part of the modeling, the stable accretion disk undergoes perturbations due to the infall of matter originating from a disrupted star or other physical mechanisms. This perturbation causes matter to fall from the outer boundary of the established stable accretion disk toward the black hole. The perturbation is induced with a constant density of $\rho = 2$ in arbitrary units, and the pressure profiles are carefully adjusted to ensure that the speed of sound is $C_s=0.1$ at the perturbation location, which is at the outer boundary of the computational domain. This meticulous adjustment ensures the appropriate conditions for the simulation, especially at the domain outer limits.

The numerical simulations are conducted on the equatorial plane, where gas is introduced from the outer boundary with radial and angular velocities $V^r = -0.01$ and $V^{\phi}= V_{\infty}\sqrt{\gamma^{\phi\phi}}\sin\phi$, respectively, for $0<\phi/rad<0.4$ at the outer boundary. $V_{\infty}$ represents the asymptotic velocity of matter at the computational domain outer boundary, and different values are used in various models as indicated in Table \ref{Inital_Con}. This configuration ensures that the gas falls towards the black hole at supersonic speeds, generating significant perturbation in the system. This setup allows for a more comprehensive analysis of its behavior and dynamics. Additional parameters used in the simulations, including the asymptotic velocity for both Kerr and Gauss-Bonnet black holes, are listed in Table \ref{Inital_Con}.

\begin{table}
\footnotesize
\caption{The initial model adopted for the numerical simulation of Kerr and Gauss-Bonnet black holes.
  $Model$, $type$, $\alpha$,  $a/M$, and  $V_{\infty}$ are the name of the model,
  the gravity, Gauss-Bonnet coupling constant, the black hole rotation parameter, and
  asymptotic velocity used in angular velocity, respectively.}
 \label{Inital_Con}
\begin{center}
  \begin{tabular}{ccccc}
    \hline
    \hline

    $Model$        & $type$        & $\alpha (M^2)$ & $a/M$ & $V_{\infty}$ \\
    \hline
    $K09$          &               & $-$       & $0.28$  & 0.2     \\    
    $K055$         & $Kerr$        & $-$       & $0.55$ & 0.2     \\
    $K09$          &               & $-$       & $0.9$  & 0.2     \\
    \hline
    $V02\_EGB1$    &               & $0.4488$  & $0.55$ & $0.2$    \\
    $V02\_EGB2$    &               & $0.3032$  & $0.55$ & $0.2$    \\
    $V02\_EGB3$    &               & $0.127$   & $0.55$ & $0.2$    \\
    $V02\_EGB4$    & $Gauss-Bonnet$& $-0.502$  & $0.55$ & $0.2$    \\
    $V02\_EGB5$    &               & $-1.659$  & $0.55$ & $0.2$    \\
    $V02\_EGB6$    &               & $-4.343$  & $0.55$ & $0.2$    \\
    $V02\_EGB7$    &               & $-5.615$  & $0.55$ & $0.2$   \\    
    \hline
    $EGB04488\_V1$ &               & $0.4488$  & $0.55$ & $0$      \\
    $EGB04488\_V2$ &               & $0.4488$  & $0.55$ & $0.01$   \\
    $EGB04488\_V3$ & $Gauss-Bonnet$& $0.4488$  & $0.55$ & $0.4$    \\
    $EGB04488\_V4$ &               & $0.4488$  & $0.55$ & $0.6$    \\
    \hline
    $EGBN4343\_V1$ &               & $-4.343$  & $0.55$ & $0$      \\
    $EGBN4343\_V2$ & $Gauss-Bonnet$& $-4.343$  & $0.55$ & $0.4$    \\
    $EGBN4343\_V3$ &               & $-4.343$  & $0.55$ & $0.6$    \\    
    \hline
    \hline
  \end{tabular}
\end{center}
\end{table}

In this paper, we have used the same number of points in both the radial and azimuthal directions as in \citet{Donmez2023arXiv231013847D}. To observe the impact of perturbation on the stable disk and ensure its transition to a steady state after perturbation, all models have been run for a minimum of $t=30000M$. For the simulations, periodic, inflow, or outflow boundary conditions have been employed to specify the necessary boundary points in the radial and azimuthal directions \citep{Donmez2023arXiv231013847D}.

In this study, we uncover the physical characteristics of $QPOs$ caused by spiral shock waves resulting from the interaction of the accretion disk with the strong gravitational field of the black hole near its horizon. The study focuses on scenarios where this interaction occurs in the equatorial plane, preserving spherical symmetry. Moreover, due to the strong gravitational force near the black hole horizon, the gravitational timescale is significantly shorter than the timescales associated with viscosity and magnetic field effects. Consequently, throughout the paper, the influence of magnetic fields and viscosity on the formation of shock waves on the disk is considered negligible, especially in the strong gravitational region. This assumption is based on the theoretical understanding that within $r<100M$, gravitational forces predominate over other physical forces.

\section{Dynamics of Perturbed Accretion Disk}
\label{dynamics}

\begin{figure*}
  \vspace{1cm}
  \center
  \psfig{file=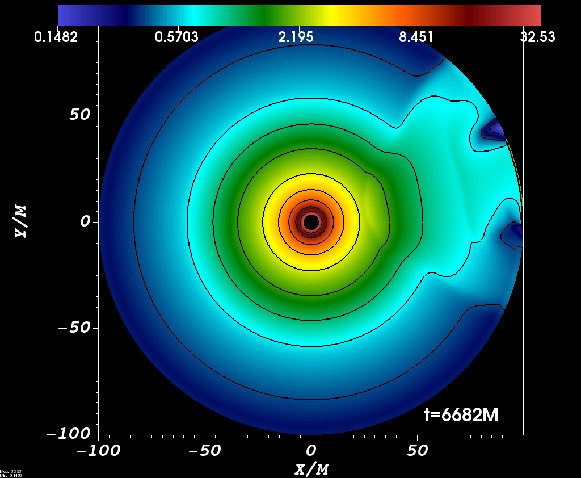,width=5cm,height=4.8cm}
  \psfig{file=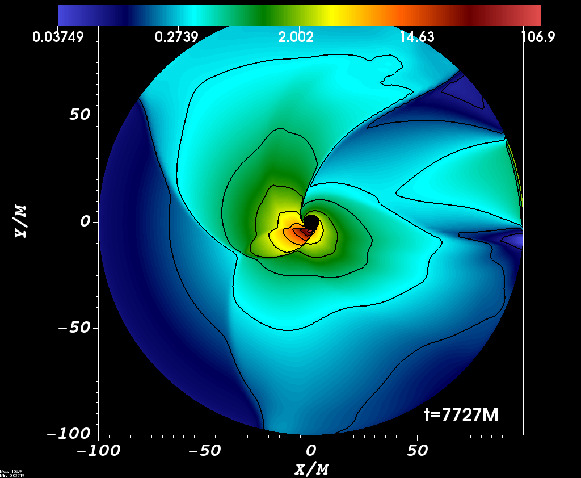,width=5cm,height=4.8cm}
  \psfig{file=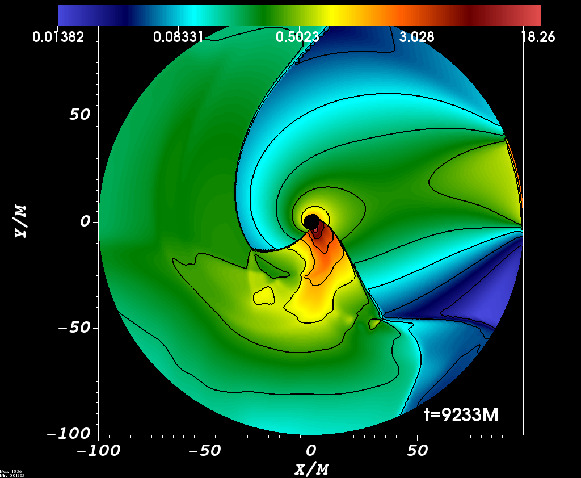,width=5cm,height=4.8cm}
  \psfig{file=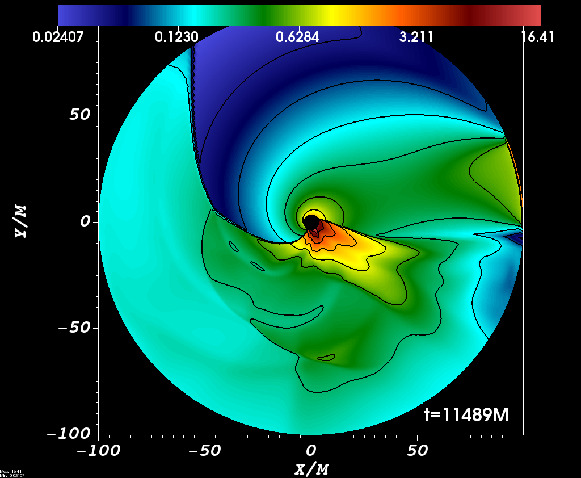,width=5cm,height=4.8cm}
  \psfig{file=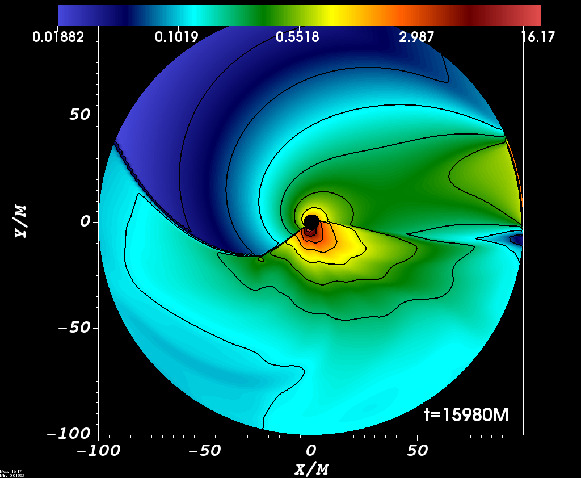,width=5cm,height=4.8cm}
  \psfig{file=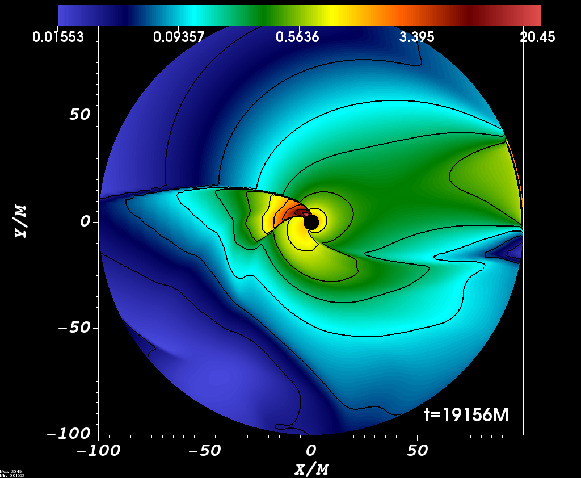,width=5cm,height=4.8cm}
  \psfig{file=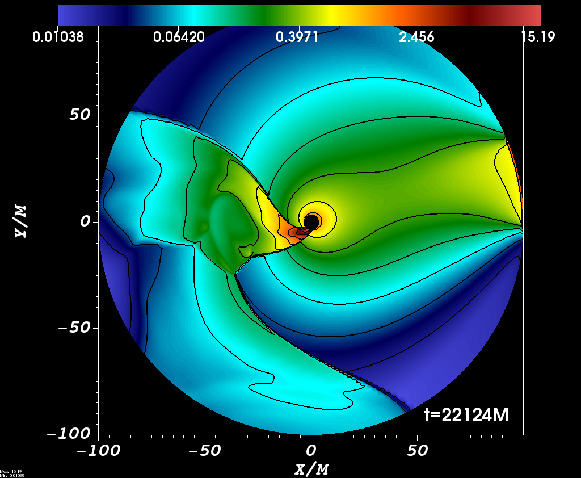,width=5cm,height=4.8cm}
  \psfig{file=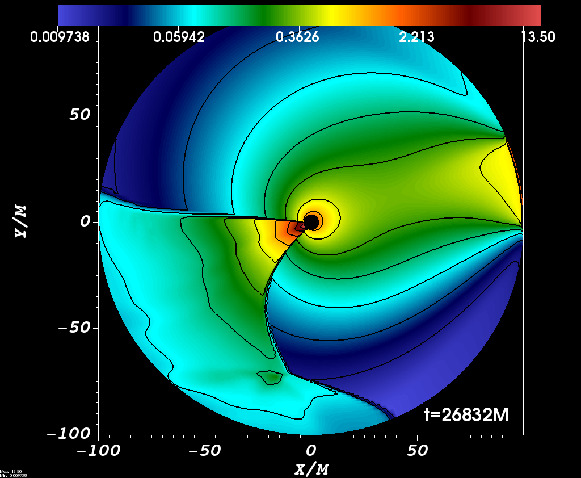,width=5cm,height=4.8cm}
  \psfig{file=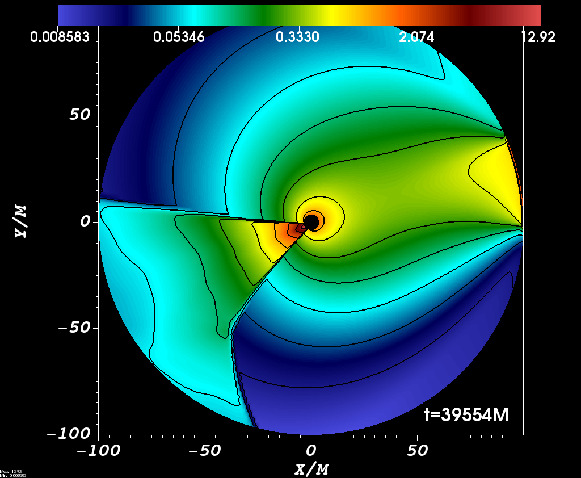,width=5cm,height=4.8cm}  
  \caption{The formation of shock waves from the perturbation of the initially stable disk, the stabilization of these shock waves, and the eventual steady-state oscillation at late times are observed. The $m=1$ and $m=2$ spiral modes have been created by perturbations from the right side of the disk, each with different angular and radial velocities.
}
\label{Color1}
\end{figure*}

\begin{figure*}
  \vspace{1cm}
  \center
  \psfig{file=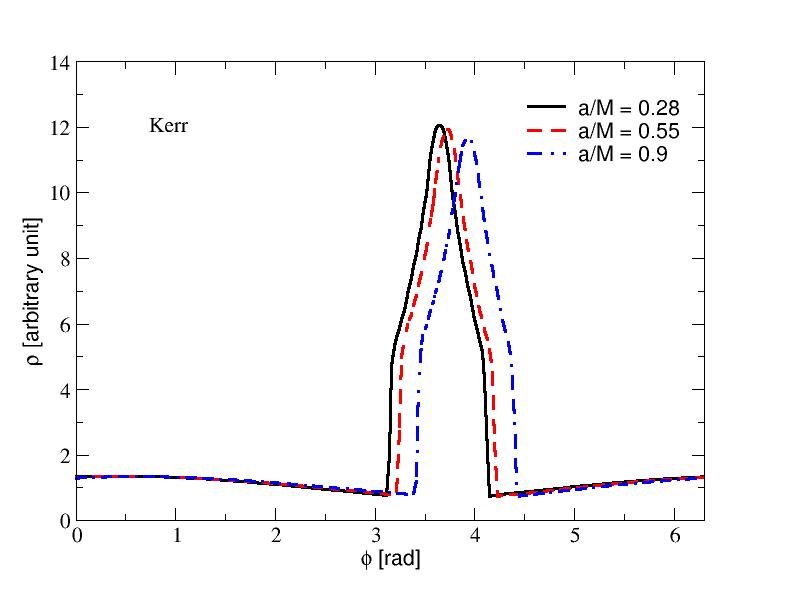,width=7.5cm,height=7cm}
   \hspace{0.7cm}  
  \psfig{file=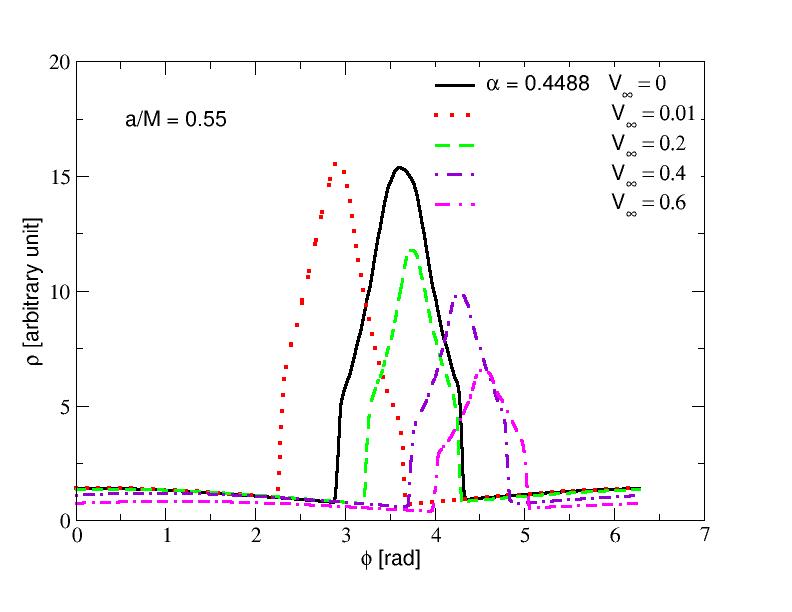,width=7.5cm,height=7cm}
  \caption{The variation of the rest-mass density along the azimuthal coordinate is depicted as a function of $a/M$ and $V_{\infty}$. On the left side of the figure, the impact of the black hole rotation parameter on the shock waves generated after the disk reaches steady state is illustrated. The graph on the right displays the density variation as a function of the asymptotic velocity for a given $a/M = 0.55$.
}
\label{dens_change}
\end{figure*}

\begin{figure*}
  \vspace{1cm}
  \center
  \psfig{file=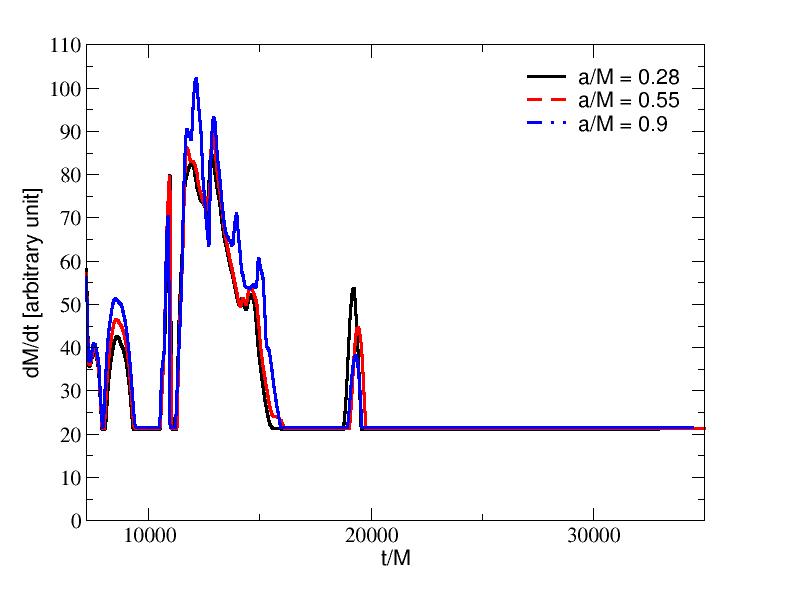,width=7.5cm,height=7cm}
   \hspace{0.7cm}  
  \psfig{file=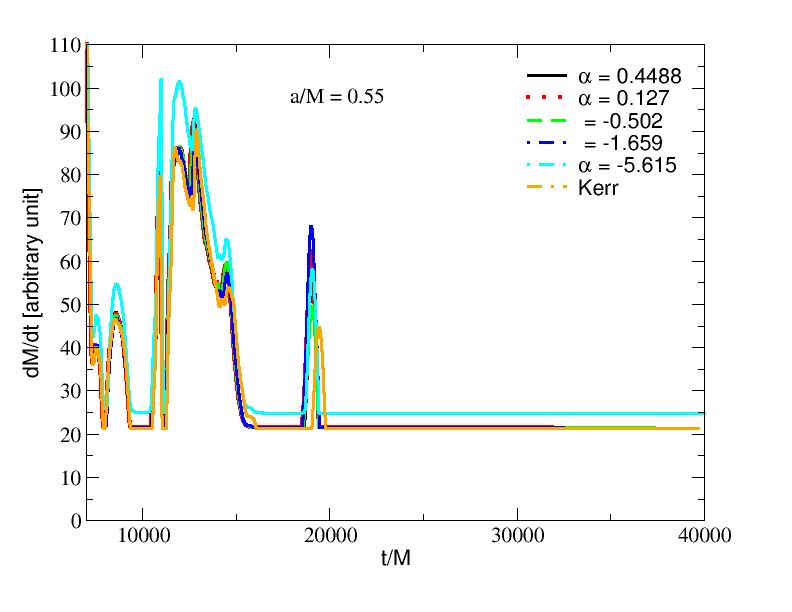,width=7.5cm,height=7cm}
  \caption{The variation of the mass accretion rate over time is depicted. In the left graph, it is shown how the mass accretion rate changes during the evolution depending on different black hole spin parameters after the disk is perturbed. In the right graph, the mass accretion rate has been numerically calculated for different EGB coupling constants ($\alpha$).
}
\label{mass_acc}
\end{figure*}

\begin{figure*}
  \vspace{1cm}
  \center
  \psfig{file=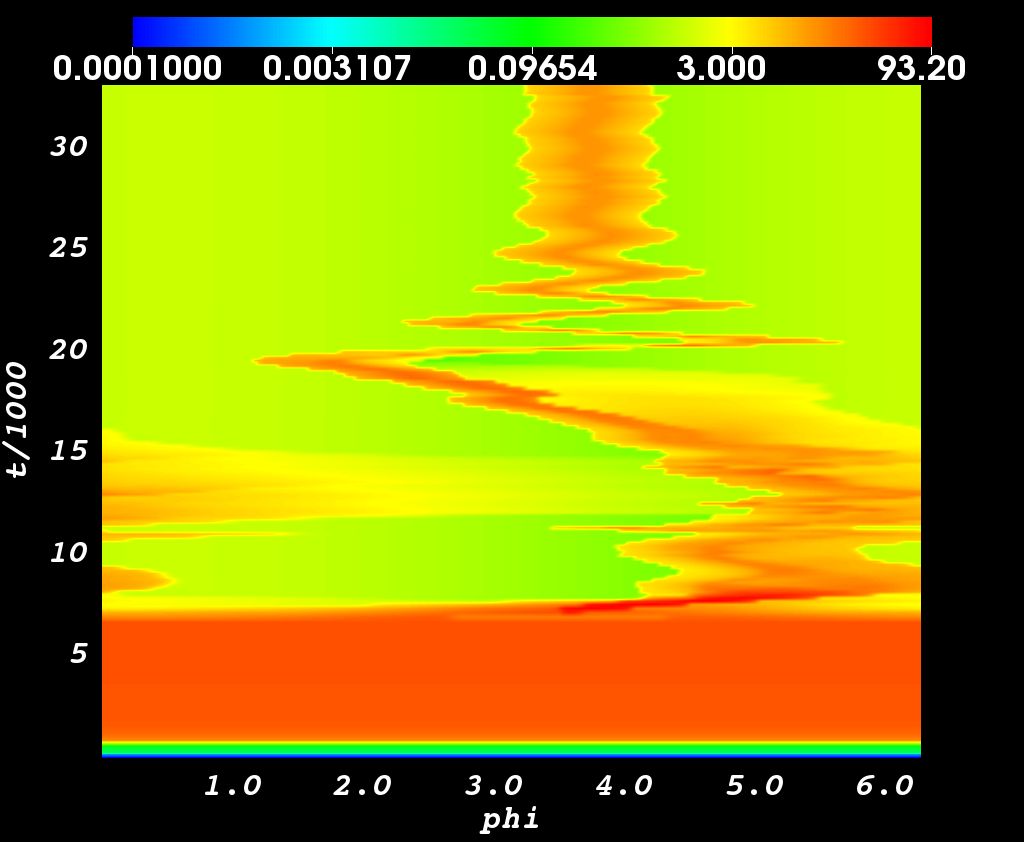,width=7.5cm,height=7cm}
  \psfig{file=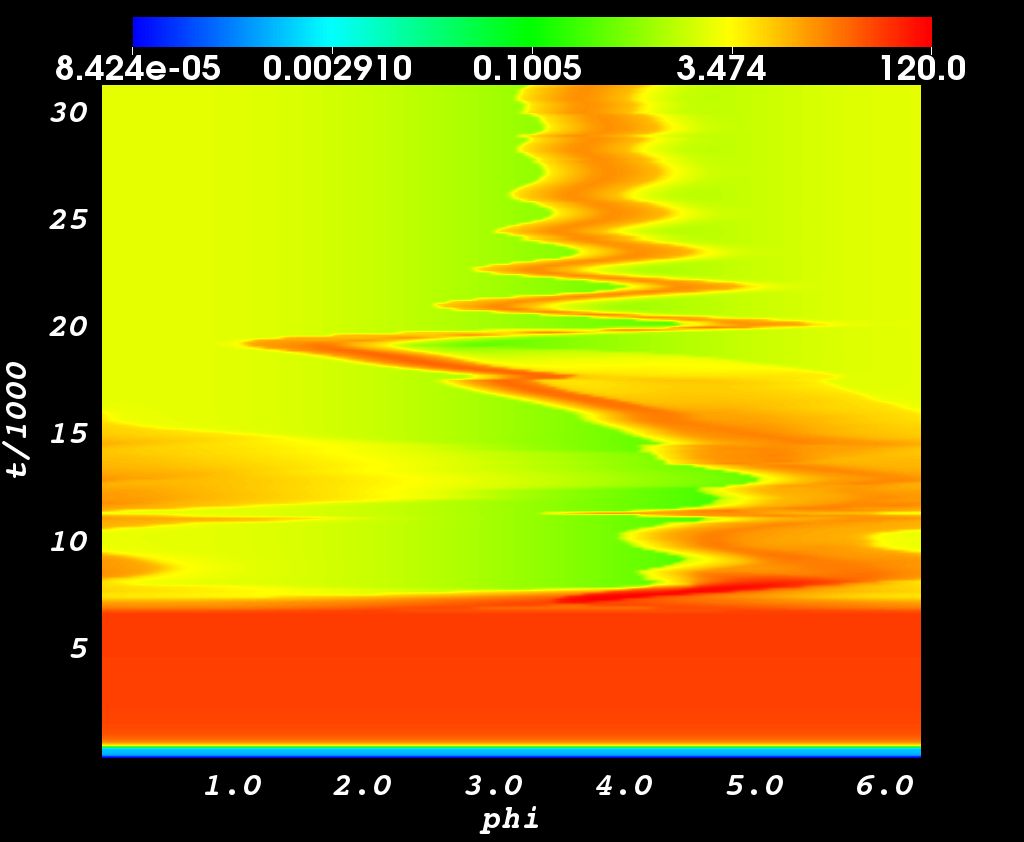,width=7.5cm,height=7cm}
  \caption{ The time-dependent change of the logarithmic density at the fixed radial point $r=3.88M$ in the azimuthal direction, $\phi/rad$. The left panel depicts the time evolution of the shock wave resulting from perturbations in the disk around the Kerr black hole (model: $K055$). Conversely, the right panel illustrates the change around the EGB black hole for $\alpha=-5.6155$ (model: $V02\_EGB7$). In both cases, $V_{\infty}=0.2$ is chosen.
}
\label{Time_Vs_ang}
\end{figure*}

\begin{figure*}
  \vspace{1cm}
  \center
  \psfig{file=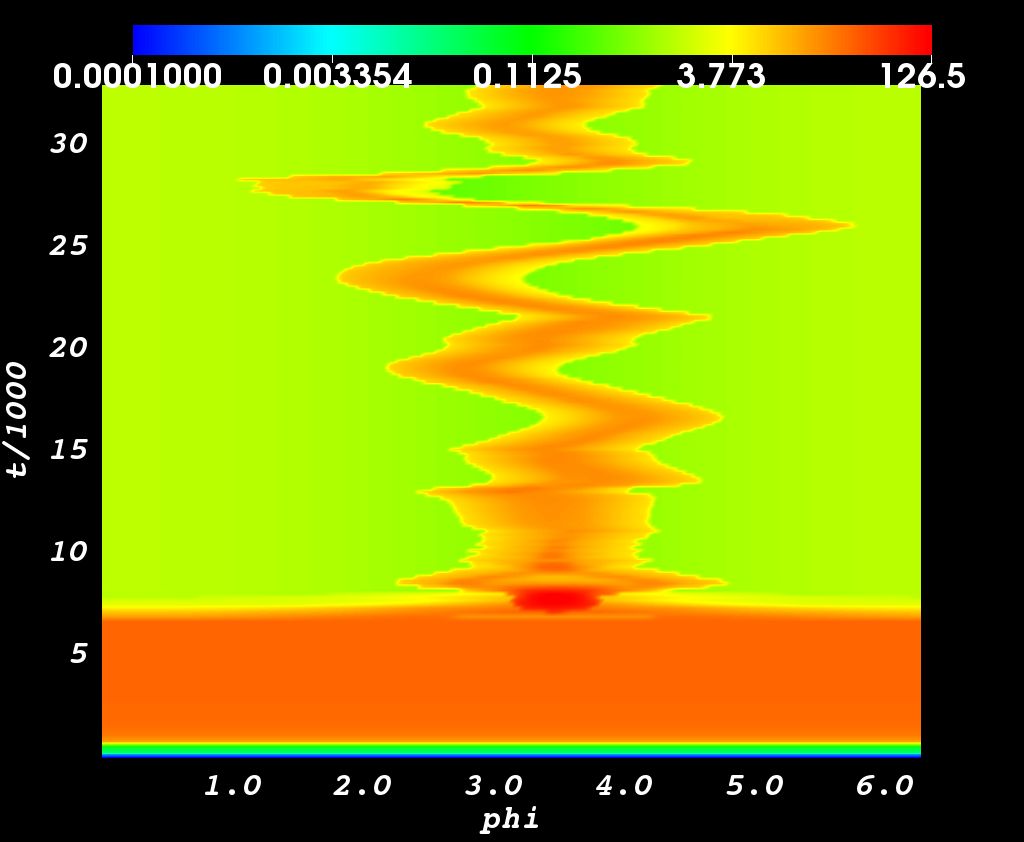,width=7.5cm,height=7cm}
  \psfig{file=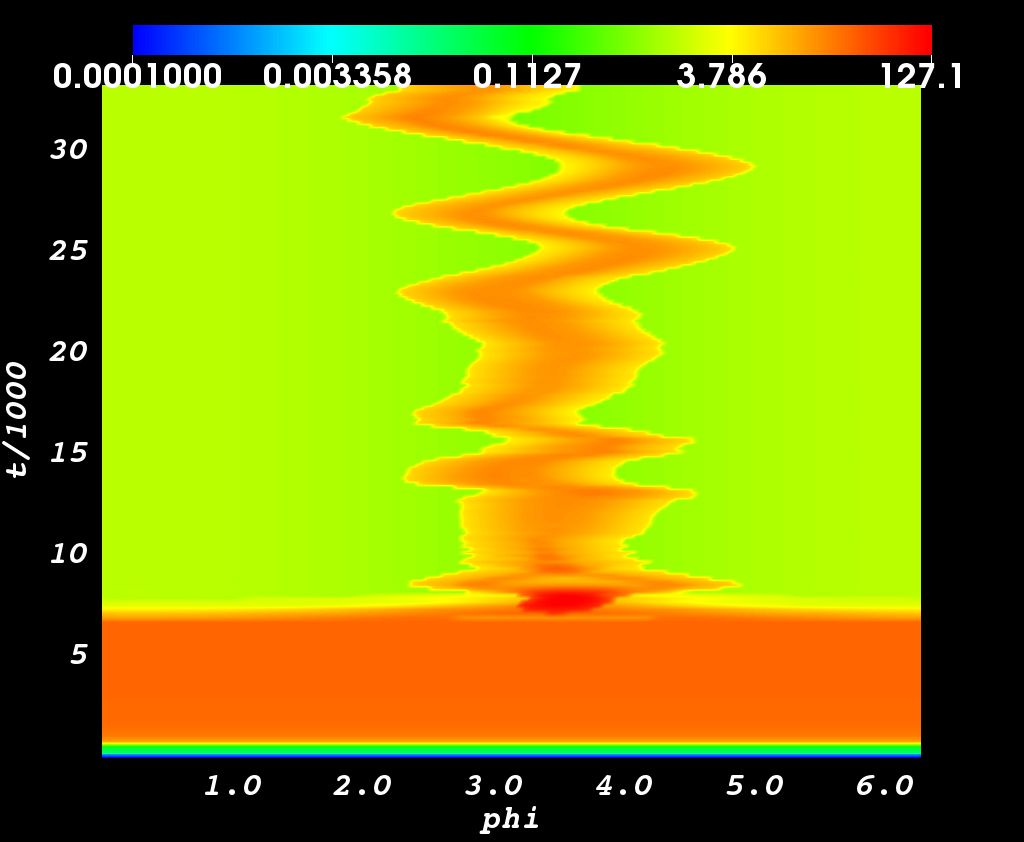,width=7.5cm,height=7cm}
  \psfig{file=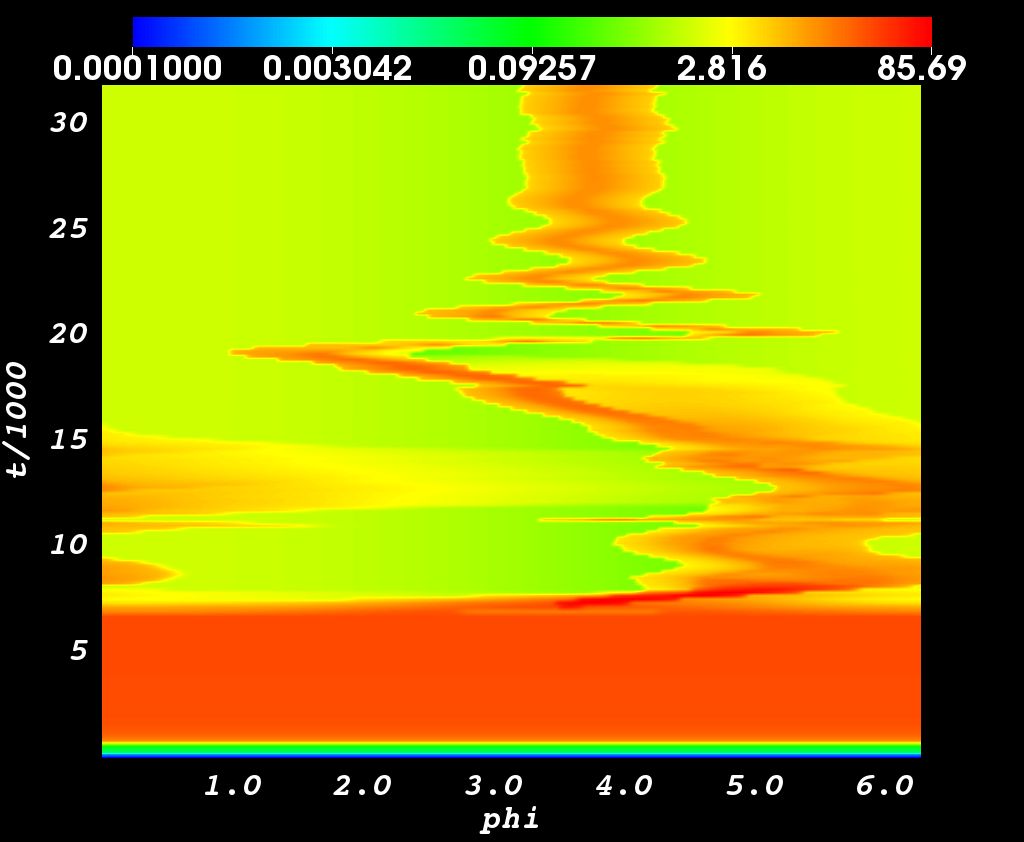,width=7.5cm,height=7cm}
  \psfig{file=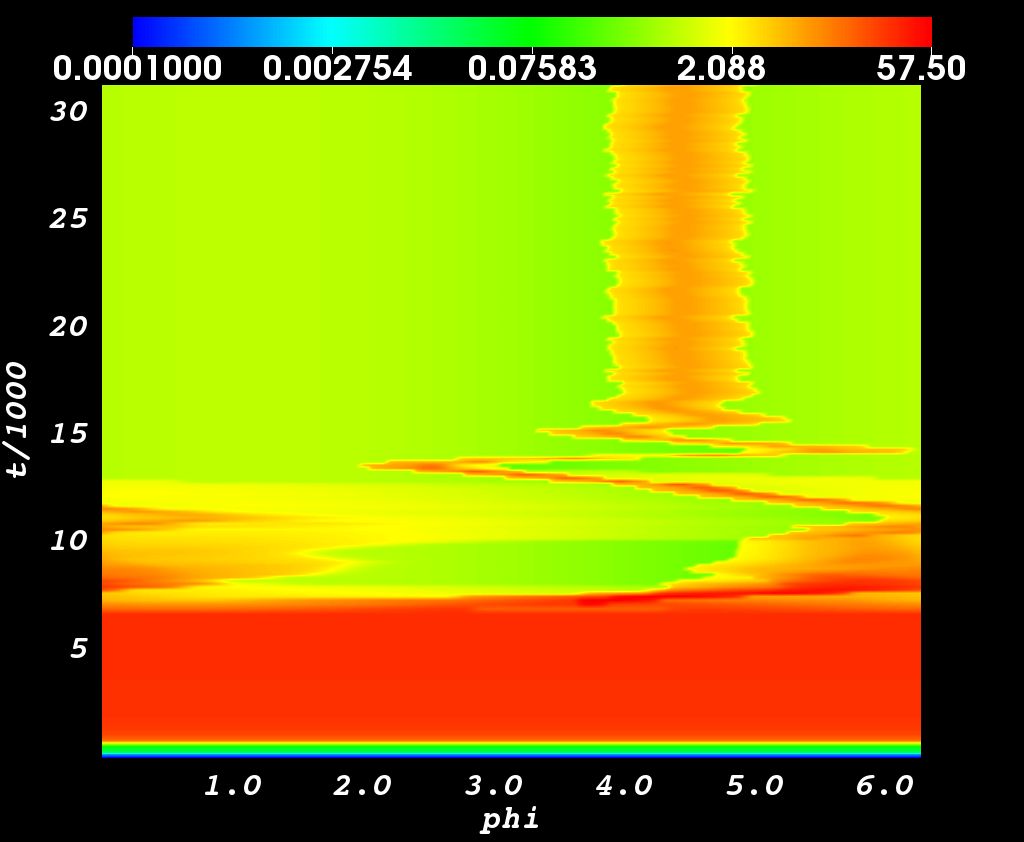,width=7.5cm,height=7cm}  
  \caption{Similar to Fig.\ref{Time_Vs_ang}, this figure pertains to other models. Here, while $\alpha=0.4488$, $V_{\infty}$ increases from top to bottom. The top two panels represent the temporal oscillation of the shock wave for the $EGB04488\_V1$ and $EGB04488\_V2$ models, while the bottom two panels represent the same for the $V02\_EGB1$ and $EGB04488\_V4$ models.
}
\label{Time_Vs_ang_M2}
\end{figure*}

\begin{figure*}
  \vspace{1cm}
  \center
  \psfig{file=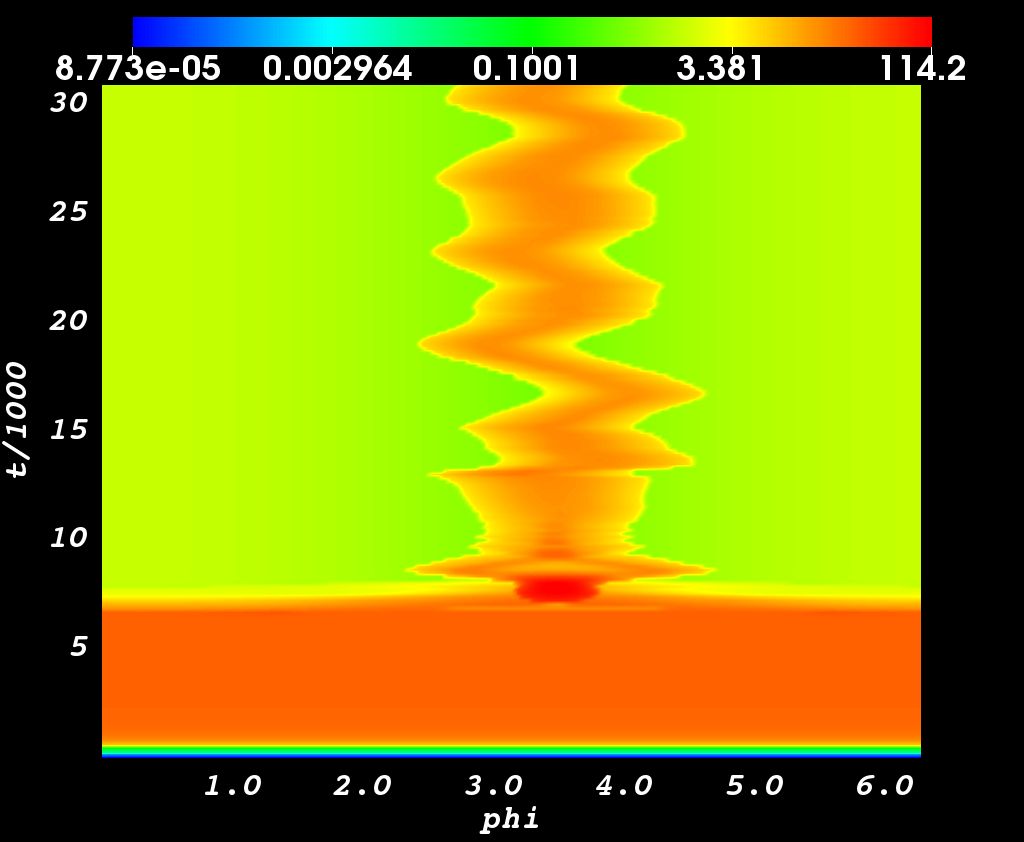,width=7.5cm,height=7cm}
  \psfig{file=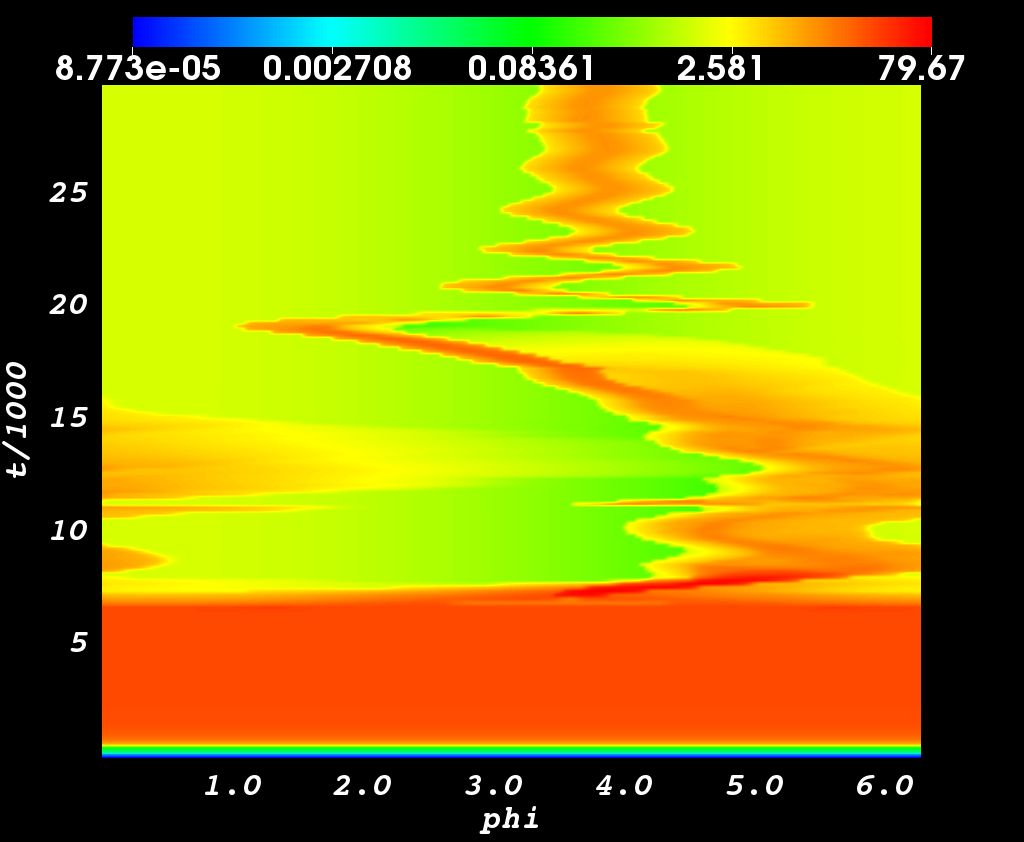,width=7.5cm,height=7cm}
  \psfig{file=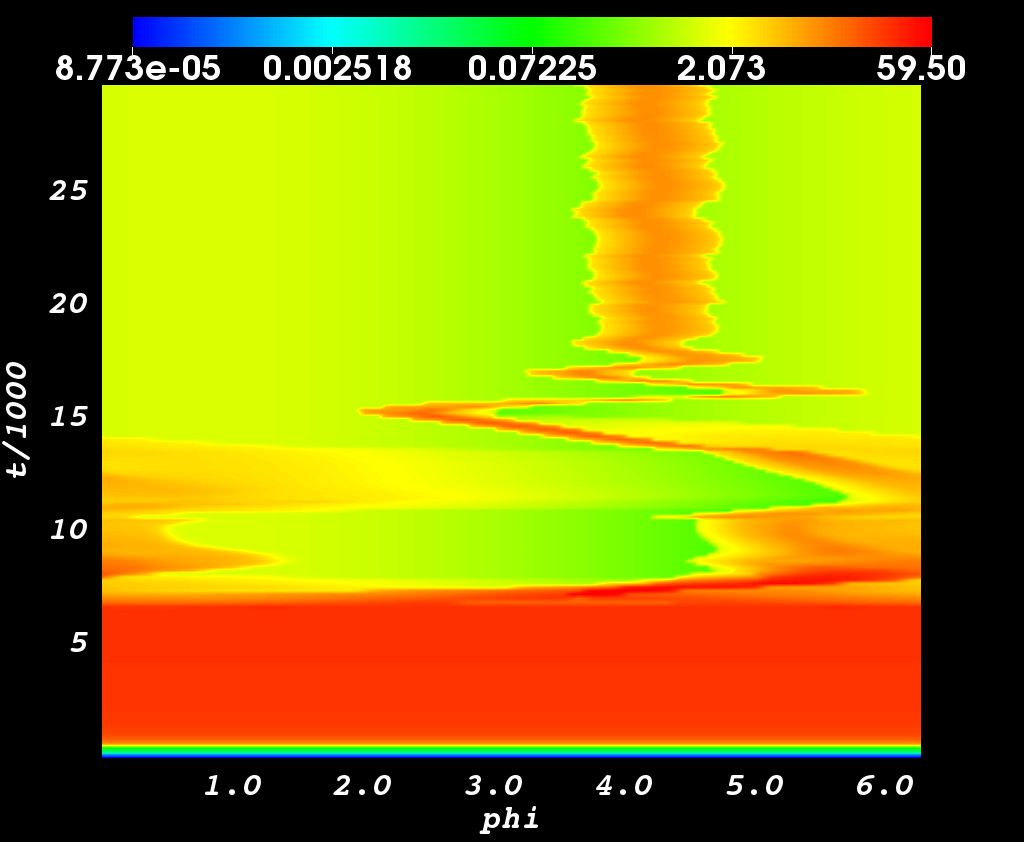,width=7.5cm,height=7cm}
  \psfig{file=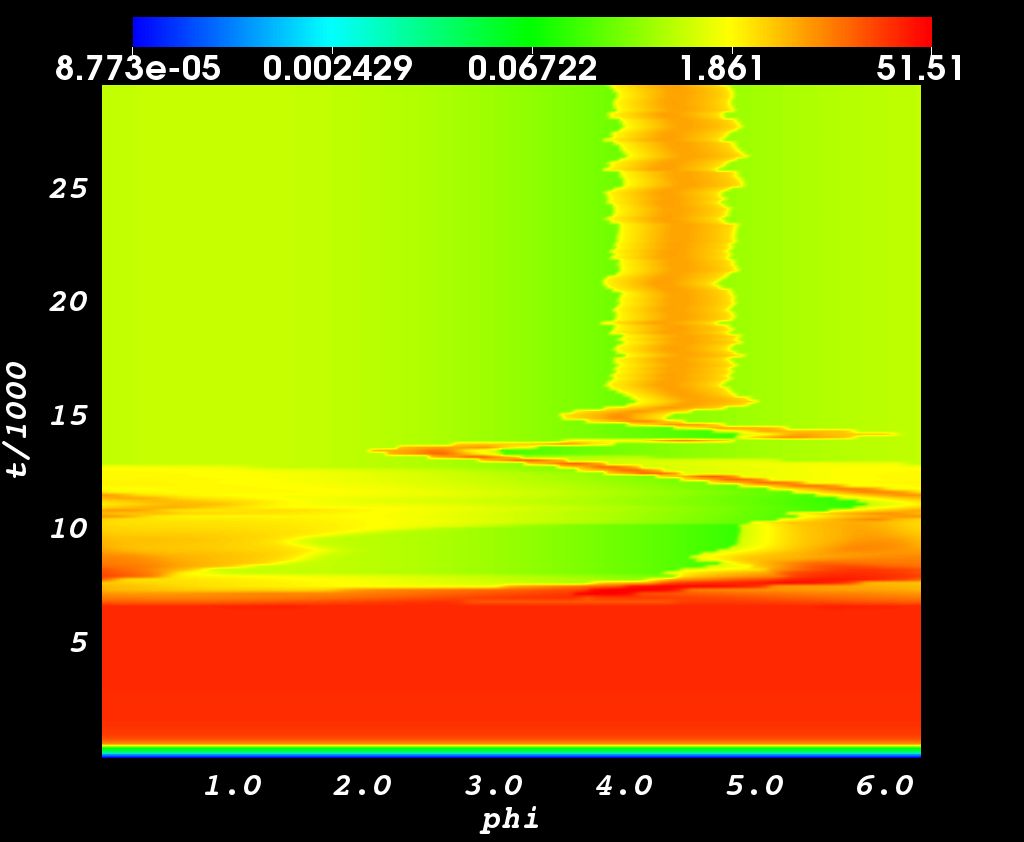,width=7.5cm,height=7cm}  
  \caption{Similar to Fig.\ref{Time_Vs_ang_M2}, this figure corresponds to $\alpha=-4.343$. The top two panels represent the temporal oscillation of the shock wave for the $EGBN4343\_V1$ and $V02\_EGB6$ models, while the bottom two panels represent the same for the $EGBN4343\_V2$ and $EGBN4343\_V3$ models.
}
\label{Time_Vs_ang_M3}
\end{figure*}

There are several ways to demonstrate the behavior of the disk around the black hole. One method is to investigate the characteristic properties of the shock waves believed to cause $QPOs$ on the disk in the strong gravitational region. This involves studying the oscillation of these shock waves, their dependence on the space-time geometry, and the physical properties of the perturbations. Figures \ref{Color1}, \ref{dens_change}, \ref{mass_acc}, \ref{Time_Vs_ang}, \ref{Time_Vs_ang_M2}, and \ref{Time_Vs_ang_M3} are provided below to illustrate the characteristic structure of the shock waves.

In Fig. \ref{Color1}, the initially stable disk, created as a result of spherical accretion, is perturbed with radial and angular velocities varying with an asymptotic speed of $V_{\infty} = 0.2$. As seen in the top left of the figure, matter has started to fall toward the black hole from a certain area on the right side of the disk. Shortly after the matter begins perturbing the disk, a spiral shock wave forms. Over time, this shock wave has been observed as one or two-armed waves, as visible in the snapshots in the middle part of the figure. As seen in the bottom row of the figure, the disk eventually reaches a steady-state phase. After the perturbed disk reaches this steady-state, the formation of a two-armed shock wave is observed, causing a $QPO$ in a certain area directly opposite the point of perturbation. This has led to the formation of $m=1$ and $m=2$ modes.

In order to gain a better understanding of the accretion disk structure around the black hole and to reveal its dependency on various parameters, we examine how the disk density depends on the black hole rotation parameter ($a/M$) and the intensity of the perturbation, determined by the asymptotic speed $V_{\infty}$ at $r=3.88M$ in the angular direction. The left part of Fig.\ref{dens_change} shows the changes in disk density and the resulting one- and two-armed shock waves as a function of $a/M$. The rotation parameter $a/M$ slightly alters the location of the shock wave formation. Additionally, because the rotation parameter warps spacetime, it causes the shock waves near the black hole horizon to bend. This effect is slightly visible in the simulations due to the inner radius of the computational domain being located at $r=3.8 M$. The choice of this inner boundary, at $r=3.8M$, is based on the EGB black hole model where the horizon extends beyond $2M$, allowing for consistent boundary comparisons across different models. The plot at the bottom further examines the changes in disk density in the angular direction at $r=3.88M$ for $\alpha=0.4488$ at different asymptotic speeds. As shown in Fig.\ref{dens_change}, $V_{\infty}$ significantly affects both the location and intensity of the spiral waves formed on the disk. An increase in asymptotic speed leads to faster matter falling into the black hole, which decreases the density. These changes significantly impact the oscillation modes around the black hole and their observability.

Fig.\ref{dens_change} also provides a comparative demonstration of the effects of $a/M$ and $\alpha$ on the spiral shock waves formed around the black hole. As seen in the left and right parts of the figure, changes in asymptotic velocity significantly affect the dynamical structure and locations of the shock waves near the black hole. However, these changes are not as pronounced across different $a/M$ values as they are with changes in $V_{\infty}$. Therefore, the moderate value of $V_{\infty}$, discussed in subsequent sections, not only triggers instability in the disk but also excites the $QPO$ frequencies.

The mass accretion rate not only describes the dynamical behavior of the accretion disk during evolution but also indicates the rate at which matter falls into the black hole. Even small oscillations in the accretion rate can lead to the formation of $QPO$ frequencies regularly produced by the modes created by spiral shock waves on the disk. Fig.\ref{mass_acc} demonstrates the variation of the mass accretion rate as a function of time after the disk has been perturbed. The right part of Fig.\ref{mass_acc} shows how this variation is dependent on $a/M$. For each $a/M$, it decreases and increases during a certain time of the mass accretion rate observed after perturbation. During this period, one- and two-armed spiral shock waves have formed. After $t=20000M$, it is seen that the disk and the generated shock waves have reached a steady-state for each $a/M$, and the $QPOs$ of the shock waves have been observed. The right part of the same figure illustrates the change in the mass accretion rate over time at different $V_{\infty}$. Similar to the graph on the left, the mass accretion rate given in the right graph has shown similar behavior. On the other hand, it is observed that $\alpha = -5.615$ behaves differently from other $\alpha$ and Kerr models. In this model, a higher rate of matter falls into the black hole. It means that the change in the mass of the central black hole is faster than in other models. However, the mass accretion behavior of the disk is similar for each $\alpha$. Again, after $t=20000M$, the small oscillations in the disk mass accretion rate have been observed. This is sufficient for the continuous production of $QPOs$ by the spiral shock waves.

Fig. \ref{Time_Vs_ang} depicts the behavior of the shock wave formed on the disk around Kerr and EGB black holes from its initial formation until it reaches a steady state. These graphs are obtained by collecting data in the azimuthal direction at $r = 3.88 M$ at each time step, revealing the behavior of the shock wave on the disk at this fixed $r$. The left figure represents the behavior of the shock wave around the Kerr black hole with $a/M = 0.55$ (model: $K055$) during the simulation time, while the right figure (model: $V02_EGB7$) shows the change in the dynamic structure of the shock wave for the same spin parameter in the case where the EGB coupling constant is $\alpha = -5.615$.

During the time interval from the beginning of the simulation until the formation of a stable disk ($t=6000M$), the same stable disk is displayed. However, after this time when perturbations are applied, the shock waves formed around the Kerr and EGB black holes exhibit different structures during their formation and, more importantly, after reaching the steady state. It is clear that the oscillation amplitude of the shock wave formed around the EGB black hole is larger than that of Kerr. As mentioned in \citet{Donmez2023arXiv231013847D} and reiterated here, we can say that for large values of the EGB coupling constant, matter behaves more chaotically. This affects the system ability to achieve a steady state and influences the frequencies of $QPOs$. On the other hand, it is evident from the color diagram given in each snapshot that the density of the matter around the EGB black hole is greater than that around the Kerr black hole. This also significantly contributes to the amplitude of $QPO$ frequencies formed around the EGB black hole. As seen in both models, the shock wave oscillates around $\frac{\phi}{\text{rad}} = 4$, which means that the $\alpha$ value does not affect the location where the shock wave forms.

To investigate the impact of physical parameters of perturbation on the shock wave formed on the disk, we modify the angular velocity of the perturbing matter. The angular velocity used in the Bondi accretion is chosen as the angular velocity for the perturbation, which is $V^{\phi} = V_{\infty}\sqrt{\gamma^{\phi\phi}}\sin\phi$. Different values of angular velocity are obtained using various asymptotic velocities as seen in Table \ref{Inital_Con}. Fig.\ref{Time_Vs_ang_M2} illustrates the time evolution of the shock wave formed on the disk for different angular velocities, with $\alpha = 0.4488$ and $a = 0.55$. As the asymptotic velocity $V_{\infty}$ increases from top to bottom in the figure, the angular velocity also increases. The matter falling onto the black hole at different angular velocities has caused significant changes in the disk structure and the shock waves formed. As shown in Fig.\ref{Time_Vs_ang_M2}, when there is no angular velocity, the shock wave forms at $\frac{\phi}{rad}\approx 3.5$ and exhibits high-amplitude oscillations throughout the numerical simulation.

On the other hand, after perturbing the disk, it begins to exhibit $QPOs$ shortly thereafter, which continue throughout the simulation. When $V_{\infty} = 0.01$, there is a significant change in the disk angular velocity, and the behavior of the shock wave changes significantly. This situation is evident in the upper-right figure. With $V_{\infty} = 0.4$, and as the angular velocity increases significantly, the shock wave initially undergoes high-amplitude nonlinear oscillations, then transitions to low-amplitude oscillations after $t = 25000M$. Additionally, the shock wave on the disk is observed to exhibit $QPO$ at $\frac{\phi}{rad}\approx 4$.
When the asymptotic velocity is $V_{\infty} = 0.6$, a high degree of chaotic behavior is clearly visible immediately after the perturbation. This results in a significant shift in the shock wave position to $\frac{\phi}{rad}\approx 5$, where it exhibits $QPOs$.

In conclusion, increasing the angular velocity not only changes the location of the shock wave on the disk but also alters its physical characteristics, leading to different $QPO$ frequencies on the disk. Furthermore, an increase in the angular velocity results in a decrease in the amount of matter on the disk, signifying a significant reduction in maximum density and causing more matter to fall into the black hole.

In Fig.\ref{Time_Vs_ang_M3}, similar to Fig.\ref{Time_Vs_ang_M2}, different asymptotic velocities are used, but this time with $\alpha = -4.343$. The effects of these velocities on the disk and shock waves are examined. The comparison reveals that the angular velocity more significantly influences the physical characteristics of the shock wave in relation to $\alpha$. However, $\alpha$, especially for large negative values, significantly affects the disk oscillations, making them more chaotic. This is clearly visible in the figure. These changes also affect the characteristics of the numerically calculated $QPOs$. In Fig.\ref{Time_Vs_ang_M3}, it is observed that the large negative values of $\alpha$ significantly influence the dynamic structure of the disk, compared to $V_{\infty}=0$ (model: $EGB04488_V1$) in Fig.\ref{Time_Vs_ang_M2}. In both cases, similar behaviors are observed immediately after the perturbation. However, in Fig.\ref{Time_Vs_ang_M2}, an increase in oscillation amplitude is observed after $t=20000M$, which is suppressed by a large $\alpha$, leading to more compact disk oscillation throughout the evolution.

\section{Instabilities}
\label{instability}

\subsection{Instabilities of $m=1$ and $m=2$ modes}
\label{m1m2}

Disturbing the stable disk around the black hole results in the formation of spiral density waves on the disk. This, in turn, gives rise to the Papaloizou-Pringle instability within the disk. When the stable disks, formed due to global accretion, are disturbed, the disk deviates from its axisymmetric structure. In other words, spiral density waves are generated in the disk. These density waves lead to perturbations in the disk angular momentum structure \citep{Donmez4, Wessel2023PhRvD}. Consequently, the Papaloizou-Pringle instability causes significant changes in the angular momentum of the matter within the disk \citep{Bugli2018MNRAS}. As a result, density waves, including $p$, $g$, and radial modes captured by these waves, give rise to the generation of $QPOs$. To determine the oscillation state of these modes, we perform the calculation of $P_m = \frac{1}{r_{out}-r_{in}}\int_{r_{in}}^{r_{out}} \ln\left(\left[\text{Re}(w_m(r))\right]^2 + \left[\text{Im}(w_m(r))\right]^2\right)dr$. Here, $r_{in}$ and $r_{out}$ are the inner and outer radii of the computational domain, respectively, and $\text{Im}(w_m(r))$ and $\text{Re}(w_m(r))$ are the imaginary and real parts of the mode, respectively \citep{Donmez4,Bugli2018MNRAS}.

\begin{figure*}
  \vspace{1.0cm}
  \center
  \psfig{file=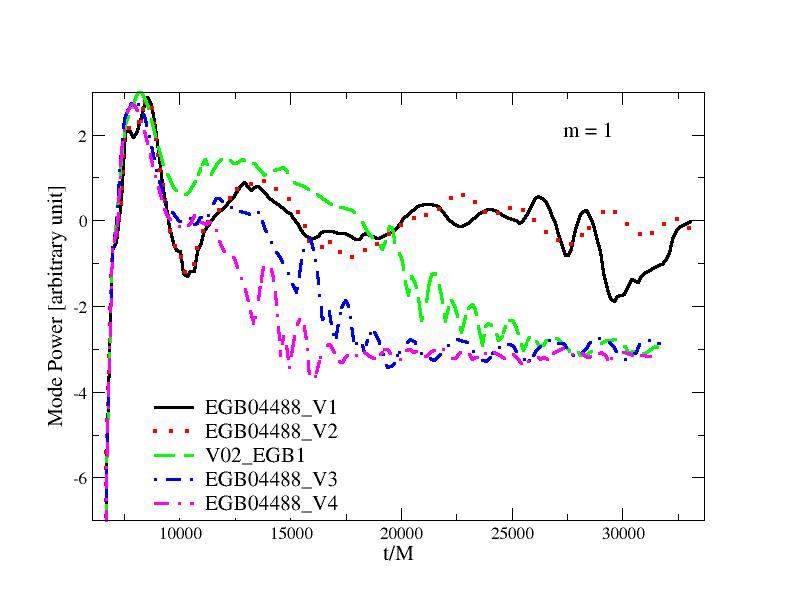,width=8cm,height=7cm}
  \hspace{0.7cm}  
  \psfig{file=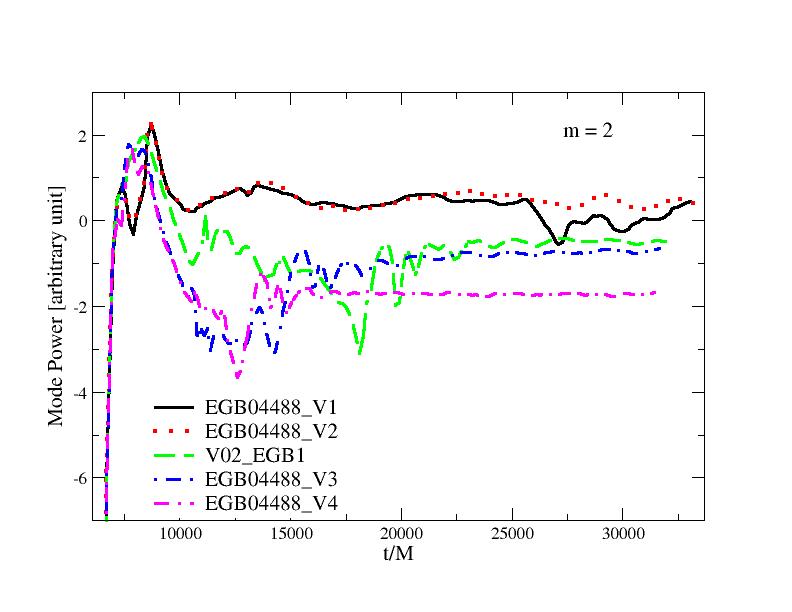,width=8cm,height=7cm}\\  
  \caption{The time-dependent variation of instability for different models in the $m=1$ and $m=2$ azimuthal modes is presented. For a consistent spin parameter ($a/M=0.55$) and EGB coupling constant ($\alpha=0.4488$), we depict the instability variation on the disk around the black hole for different perturbation velocities in the models $EGB04488\_V1$ ($V_{\infty}=0$), $EGB04488\_V2$ ($V_{\infty}=0.01$), $V02\_EGB6$ ($V_{\infty}=0.2$), $EGB04488\_V3$ ($V_{\infty}=0.4$), and $EGB04488\_V4$ ($V_{\infty}=0.6$).
} 
\label{Mode_Power1}
\end{figure*}

The Papaloizou-Pringle instability that forms on the disk leads to the growth of the $m=1$ mode, representing a one-armed spiral wave, and $m=2$ modes, representing two-armed spiral waves. This behavior is illustrated in Fig.\ref{Mode_Power1}. These modes grow in conjunction with perturbations on the disk and reach saturation around $t \approx 8500M$. As seen in Fig.\ref{Mode_Power1}, the $m=1$ and $m=2$ modes exhibit different behaviors. This behavior is not only dependent on the black hole spin and the EGB coupling constant, as shown in our previous works \citep{Donmez4,Donmez2023arXiv231013847D}, but it is also strongly tied to the angular velocity of the perturbation. After reaching saturation at $V_{\infty}=0$, the $m=1$ and $m=2$ modes quickly begin exhibiting $QPOs$. However, the amplitudes of these oscillations are significant, leading to the immediate appearance of $QPO$ frequencies on the disk after the perturbation. On the other hand, for $V_{\infty}=0.01$ and larger values of $V_{\infty}$, the $m=1$ and $m=2$ modes reach a quasi-periodic behavior. The archiving stability occurs later in the $m=1$ mode, potentially resulting in the dominance of the two-armed spiral wave in the $QPO$ frequencies. As $V_{\infty}$ approaches $0.6$, the transition to stability occurs rapidly after reaching saturation. The behavior of modes transitioning from saturation to instability and back to stability can explain the different behaviors observed in the $QPOs$ calculated in numerical modeling discussed later. As seen in Fig.\ref{Mode_Power1}, at $V_{\infty}=0.6$, matter is rapidly pushed inwards or outwards from the black hole, reaching a steady state after saturation and starting to exhibit $QPOs$ after some time. However, the amplitude of oscillation is lower compared to other models, primarily due to the lowest density observed in the disk around the black hole in this model.

\subsection{Root Mean Square (RMS)}
\label{rms}

The linear RMS-flux relationship for cataclysmic variables in binary systems and AGNs was first introduced by \citet{Scaringi2012MNRAS}. This relationship supports the existence of a common, general physical mechanism explaining the wide-band radiation variations in $X$-ray binaries and AGNs, independent of the source, mass, and size. RMS quantifies the deviation of mass accretion from its time-averaged value, as determined by numerical simulations, thus shedding light on the strength of disk oscillations near the black hole horizon ($r=3.88M$), where gravitational forces are intense.

\begin{figure*}
  \vspace{1.2cm}
  \center
  \psfig{file=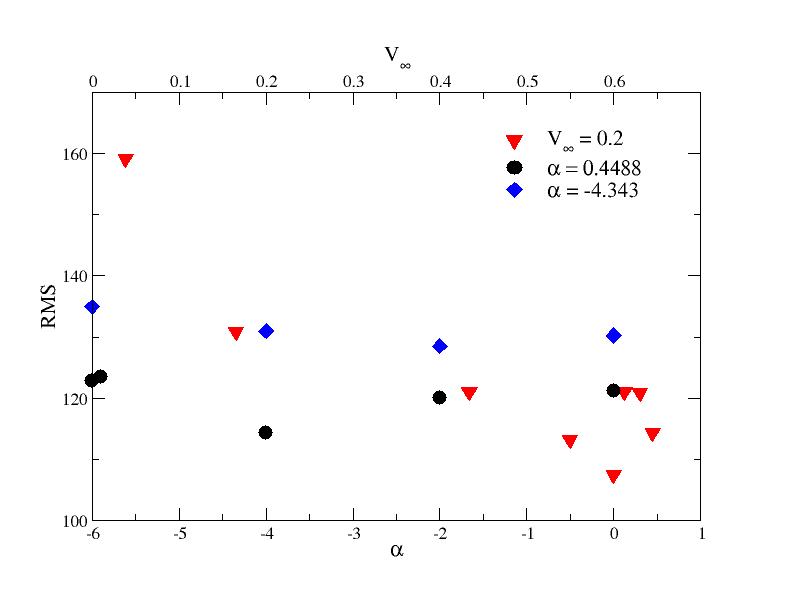,width=10cm,height=8cm}
  \caption{The variation of the RMS as a function of $\alpha$ and $V_{\infty}$ is presented. Inverted triangles represent the change in RMS with $\alpha$ for $V_{\infty} = 0.2$, while filled circles and diamonds indicate the change in RMS with $V_{\infty}$ for $\alpha = 0.4488$ and $\alpha = -4.343$, respectively. The RMS was calculated using the mass accretion rate at $r=3.88M$.
}
\label{Figrms}
\end{figure*}

Fig. \ref{Figrms} illustrates the variation in RMS for different $\alpha$ and $V_{\infty}$ scenarios, highlighting how disk oscillations and potential $QPO$ frequencies due to accretion vary with these parameters around the black hole.
Numerical calculations show that RMS is influenced by the black hole spin ($a/M$), EGB coupling constant ($\alpha$), and asymptotic velocity ($V_{\infty}$), affecting the numerically observed $QPOs$. High RMS values indicate significant amplitudes of $QPOs$ generated near the black hole, enhancing their detectability. However, models with high RMS values often exhibit more chaotic oscillations. For instance, in Fig. \ref{Figrms}, RMS is notably higher for $V_{\infty} = 0.2$ with $\alpha = -4.343$ and $\alpha = -5.615$. Consequently, the $QPO$ frequencies, other than the primary genuine mode, differ from those with relatively lower RMS values. Typically, $QPOs$ on the disk are formed through nonlinear combinations, apart from a few genuine modes. Higher RMS values signify increased disk instability, leading to frequency shifts towards higher values and generating frequencies with larger amplitudes, as observed in the top and bottom left panels of Fig. \ref{QPO_diff_alpha}.

\section{QPOs in the Perturbed Disk}
\label{QPO}

After perturbation, the disk remained unstable for a long period due to the interaction between the black hole, disk, and perturbation, leading to significant mass loss. This mass was either expelled from the computational domain or absorbed by the black hole. While matter falling into the black hole incrementally increases its mass, this mass change has been overlooked, and the black hole mass is assumed to remain constant over time. Initially, a one-armed shock wave formed on the disk, followed by a two-armed shock wave. The disk reached a steady state around $t=20000M$, with the initial instability and the shock waves causing oscillations. After returning to the steady state, the disk continued oscillating at a consistent frequency.

In numerical relativity, accurately capturing the physical nature of the $QPO$ frequencies necessitates sufficient frequency resolution. The oscillation frequencies were derived by applying the Fast Fourier Transform to the accretion rate. The mass accretion rate at $r=3.88M$, near the black hole horizon, was calculated, where the time for one orbital cycle around the black hole is approximately $48M$. Post reaching a steady state, the disk’s behavior over an additional $20000M$ was analyzed, indicating that matter at $r=3.88M$ completes about 416 orbits in a semi-stable manner, affirming the physical nature of the $QPO$ frequencies on the disk.

To uncover these $QPOs$, the mass accretion rate at the closest point to the black hole, at $r=3.88M$, is calculated. This rate is computed using the formula $\frac{dM}{dt} = \int_0^{2\pi} \alpha \sqrt{\gamma} D \left( v^r - \frac{\beta^r}{\alpha} \right) d\phi$, which is then used for power spectrum analysis. In the resulting Power Spectrum Density (PSD) graphs, the frequency axis unit is converted from geometric units to Hz using the expression $f(\text{Hz}) = f(M) \times 2.03 \times 10^5 \times \left(\frac{M_\odot}{M}\right)$. Because geometric units depend on the black hole mass, $M = 10M_{\odot}$ is assumed for all graphs. However, adjustments are made in the narrative and explanations when the mass of the black hole at the center of the source significantly deviates from $M = 10M_{\odot}$.

\begin{figure*}
  \vspace{1.2cm}
  \center
  \psfig{file=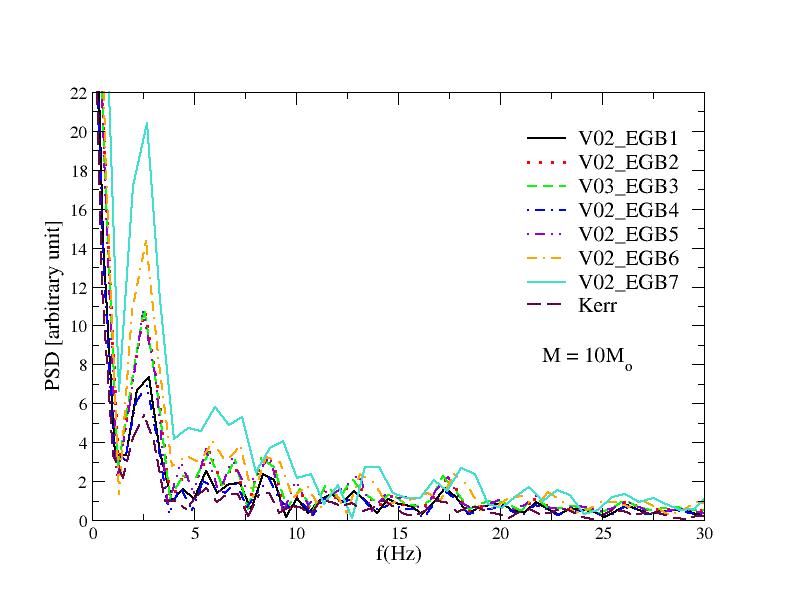,width=8cm,height=7cm}
 \hspace{0.5cm}  
 \psfig{file=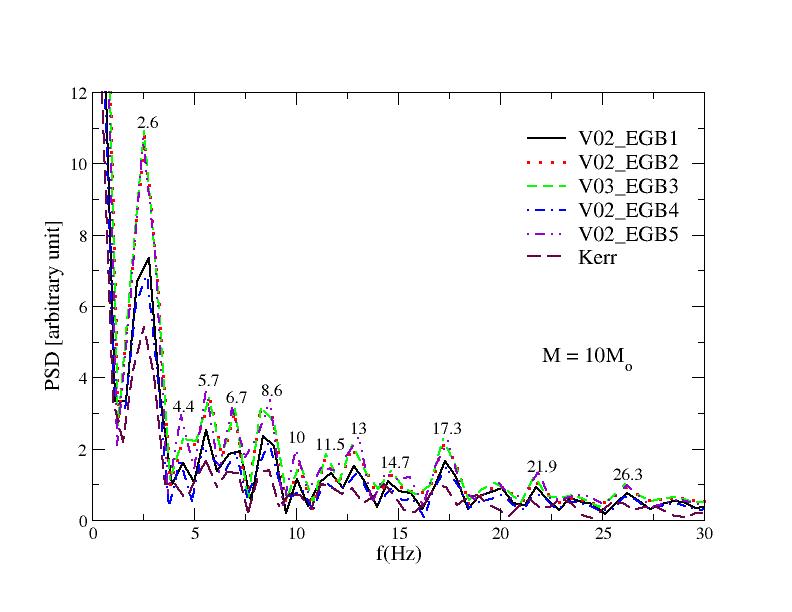,width=8cm,height=7cm}\\
   \vspace{0.5cm}
  \psfig{file=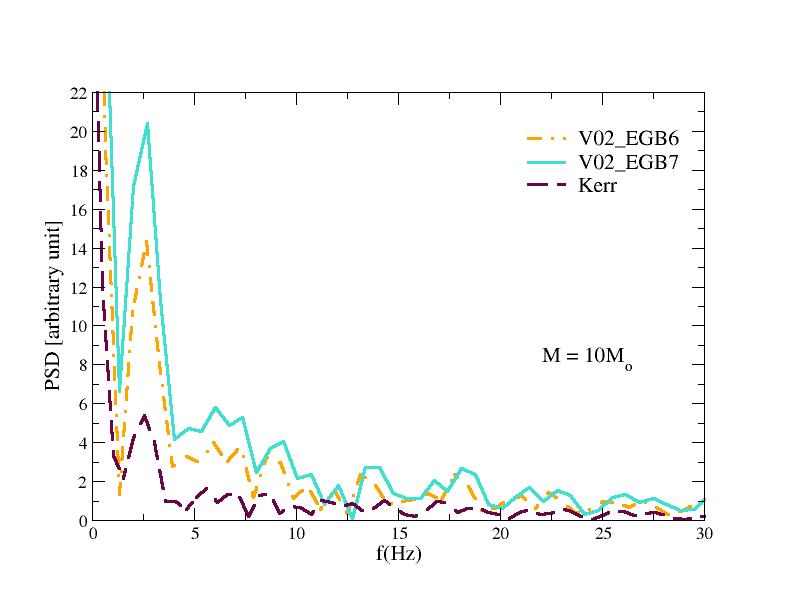,width=8cm,height=7cm}
 \hspace{0.5cm}  
  \psfig{file=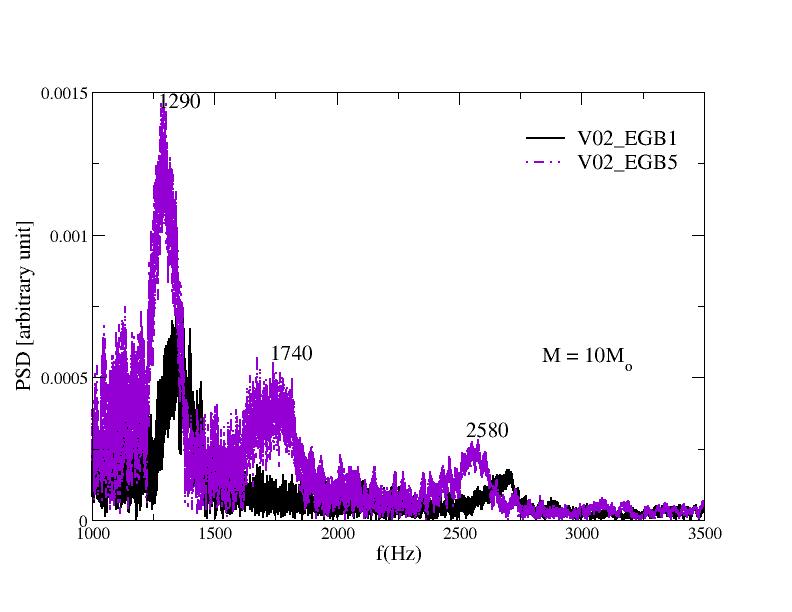,width=8cm,height=7cm} 
  \caption{Power spectrum analysis at different EGB coupling constants ($\alpha$) for $V_{\infty}=0.2$ is presented. Various panels show different data sets, highlighting the dependence of $QPOs$ generated on the disk, particularly on the angular velocity of the perturbation. The graph in the lower right panel displays oscillations at different frequencies, which are visible only in a specific data set.
}
\label{QPO_diff_alpha}
\end{figure*}

For $V_{\infty} = 0.2$, we explored how different $EGB$ constants affect the $QPO$ frequencies, comparing them with those around the Kerr black hole, as shown in Fig. \ref{QPO_diff_alpha}. The top-left graph in Fig. \ref{QPO_diff_alpha} presents a power spectrum analysis for all models at $V_{\infty} = 0.2$, focusing on the $EGB$ constants and their resultant $QPO$
 frequencies. The first genuine mode appeared at nearly consistent frequencies, but distinct differences in the second and subsequent modes, especially for large negative $\alpha$ values, were observed. These differences are highlighted in the top-right and bottom-left graphs of Fig. \ref{QPO_diff_alpha}, particularly for $\alpha = -4.343$ and $\alpha = -5.615$. Comparisons with the Kerr model illustrated how the $QPO$ frequencies diverged. The first genuine mode, emerging shortly after a one-armed spiral shock wave induced by perturbation, suppressed other modes, leading to a consistent frequency across each model. However, large negative $\alpha$ values significantly altered the $QPO$ frequencies. The second-row left graph in Fig. \ref{QPO_diff_alpha} shows that, for large negative $\alpha$, the oscillation amplitude is substantially higher than for other $\alpha$ values, suggesting greater visibility for these frequencies. In fact, the $EGB$ models consistently demonstrated higher frequency amplitudes than the Kerr model under identical conditions, indicating a potential for higher visibility in $EGB$ models at the same frequency. The first-row left graph in Fig. \ref{QPO_diff_alpha} indicated that the oscillation amplitudes and resultant $QPO$ frequencies for positive and most acceptable negative $\alpha$ values (>$-4.343$) are nearly identical. The first two frequencies are genuine modes, while subsequent frequencies are nonlinear modes that overlap. These modes persist as long as the disk oscillates, contributing to the observed $1:2:3...$ ratios. Lastly, high-frequency oscillation behaviors in the models from Fig. \ref{QPO_diff_alpha} were analyzed, leading to the second-row right graph. Models for $\alpha = 0.4488$ and $\alpha = -1.659$ generated high-frequency $QPOs$ aligning with observations, not seen in other $\alpha$ values. This suggests that these specific $\alpha$ models could explain various frequencies for $V_{\infty} = 0.2$.

\begin{figure*}
  \vspace{1cm}
  \center
  \psfig{file=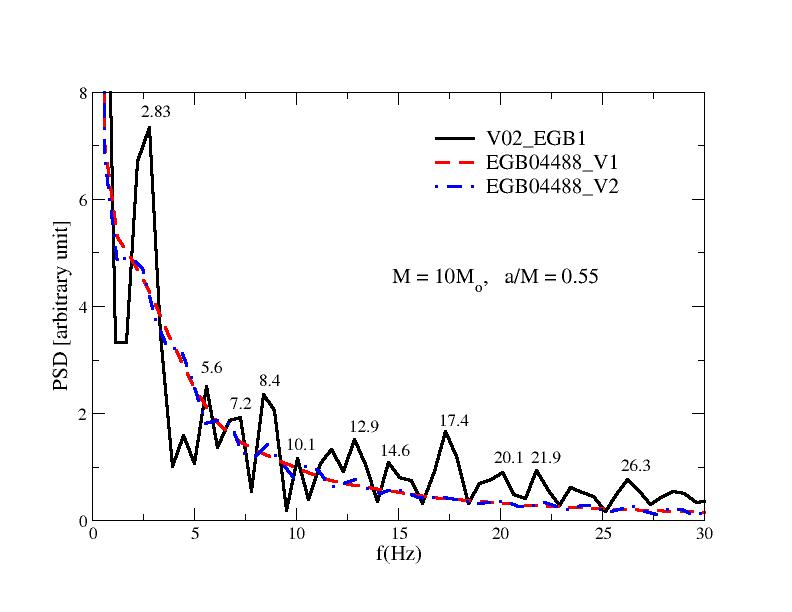,width=10cm,height=6cm} \\
  \vspace{1.0cm}
  \psfig{file=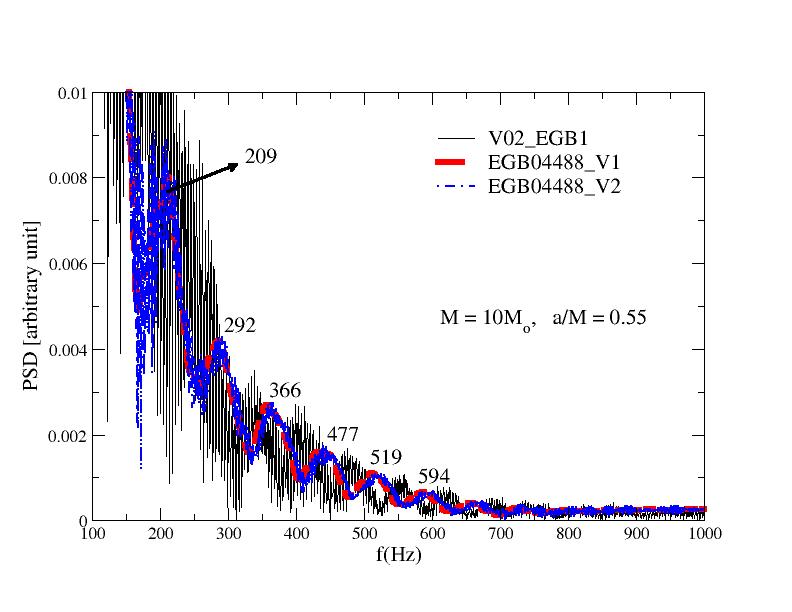,width=10cm,height=6cm}\\
  \vspace{1.0cm}
  \psfig{file=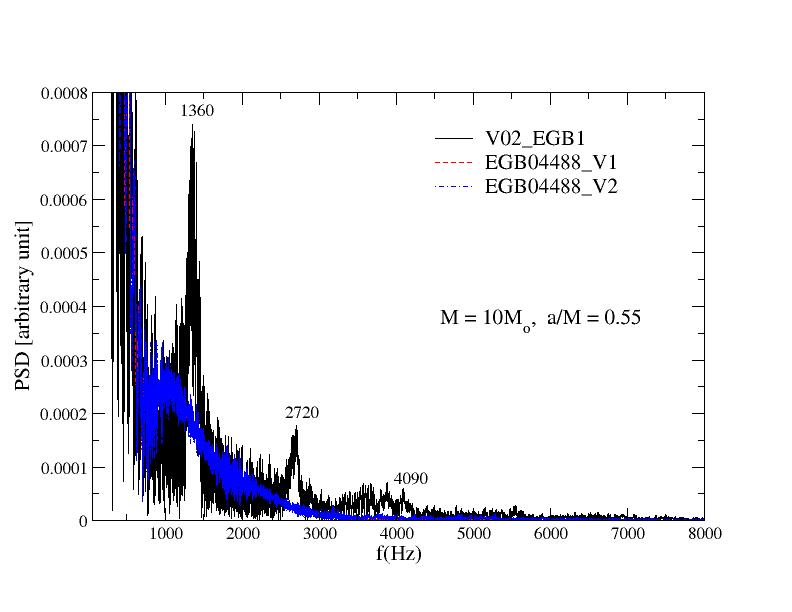,width=10cm,height=6cm}  
  \caption{The frequencies of $QPOs$ generated on the disk are observed when the perturbation has either zero or moderate angular velocities. Both low and high velocities can explain the observational results obtained from various sources.
}
\label{QPO_compare1}
\end{figure*}

Unlike Fig. \ref{QPO_diff_alpha}, Fig. \ref{QPO_compare1} models the $QPO$ behavior of oscillations on the disk at different angular velocities of perturbation for the same $EGB$ constant, $\alpha=0.4488$. These angular velocities correspond to asymptotic speed values of $V_{\infty}=0$, $0.01$, and $0.2$. As shown in the top graph of Fig. \ref{QPO_compare1}, $QPO$ oscillation frequencies do not occur at zero or low angular velocities of the perturbation, even though the same type of two-armed spiral wave is formed (refer to the graph in the upper right column of Fig. \ref{Time_Vs_ang_M2}). $QPO$ oscillations only occur for $V_{\infty}=0.2$, indicating that the angular velocity of the matter perturbing the disk significantly impacts the resulting $QPO$ behavior. We believe that the $QPO$ frequencies obtained for $V_{\infty}=0.2$ align with observed low-frequency sources, which is elaborated on in the subsequent section.

Furthermore, the oscillations at high frequencies in these three models are examined, resulting in the middle part of the graph in Fig. \ref{QPO_compare1}. While a clear conclusion is elusive, it appears that low angular velocity perturbations can explain high-frequency oscillations on the disk. At high frequencies, the oscillation frequencies of the disk, and thus the spiral wave, are in harmony for $V_{\infty}=0$ and $V_{\infty}=0.01$. We surmise that these two models may not generate any modes at low frequencies, but some predicted theoretical modes may have been captured due to the regular oscillation of the two-armed spiral wave on the disk, leading to these $QPO$ frequencies. Thus, the angular velocity at $V_{\infty}=0.2$ is seen to produce $QPO$ oscillation frequencies consistent with observations, as indicated in our previous discussions and here. Additionally, $V_{\infty}=0.2$ is observed to generate very high-frequency oscillations, as shown in the bottom graph of Fig. \ref{QPO_compare1}, which could explain phenomena observed in high-frequency $X$-ray binary systems \citep{Pasham2015ApJ}.

In Figs. \ref{QPO_diff_alpha} and \ref{QPO_compare1}, numerical models show that very high frequencies at $V_{\infty}=0.2$ are observable only at certain $\alpha$ values. One instance is illustrated in the bottom right corner of Fig. \ref{QPO_diff_alpha}, where, under specific initial conditions for $M = 10M_{\odot}$, frequencies of $1290$, $1740$, and $2580 Hz$ have been recorded, with the $1290:2580$ ratio closely matching the well-known observational $1:2$ ratio. Similarly, the bottom graph of Fig. \ref{QPO_compare1} demonstrates that numerical models can produce very high frequencies, with an observed ratio of $1330:2720:4090 \equiv 1:2:3$. Although a source with such high frequencies has not yet been observed, it is believed that these numerical results could contribute to the literature, aiding in the understanding of the physical mechanisms behind such high frequencies that might be observed in the future.

\begin{figure*}
  \vspace{1.2cm}
  \center
  \psfig{file=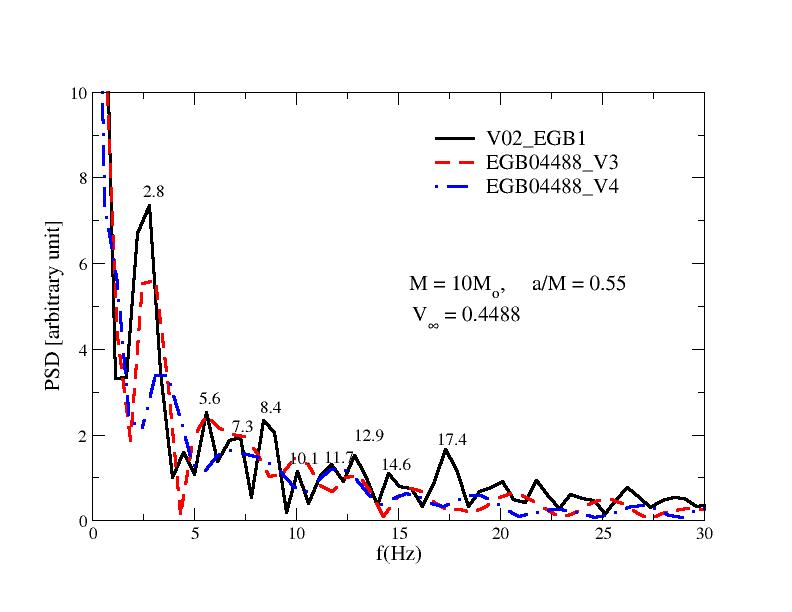,width=8cm,height=7cm}
 \hspace{0.5cm}  
 \psfig{file=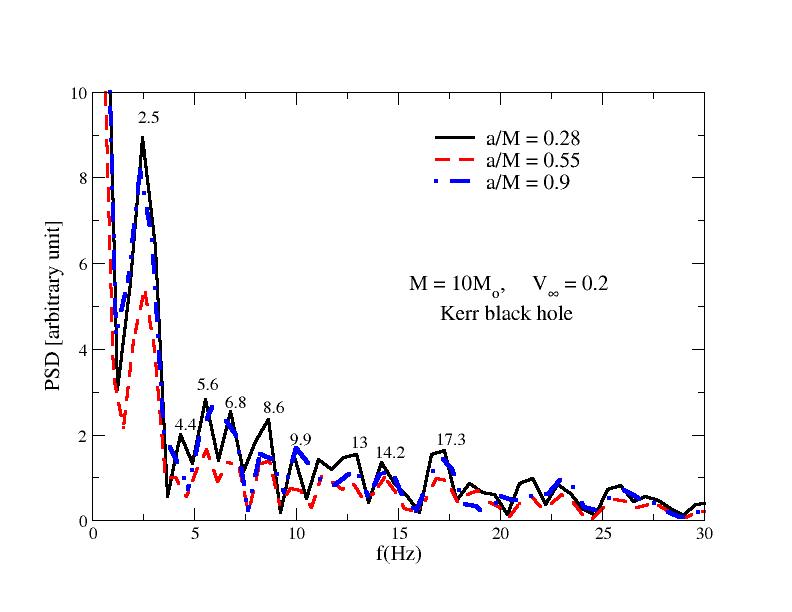,width=8cm,height=7cm}
 \caption{Left panel: The $QPO$ frequency generated on the disk due to a perturbation with high angular velocity when $a/M = 0.55$. Right panel: $QPO$ frequencies at the same angular velocity but with varying black hole spin parameters.
}
\label{QPO_diff_alpha_diff_a}
\end{figure*}

In Fig. \ref{QPO_diff_alpha_diff_a}, the left graph shows the effect of high angular velocities on the $QPOs$ around the black hole disk, while the right graph discusses the impact of the black hole spin parameter on $QPO$ formation. In Fig. \ref{Time_Vs_ang_M2}, the two-armed spiral wave and its dynamic structure around the black hole are examined at high angular velocities, specifically for $V_{\infty}=0.4$ and $V_{\infty}=0.6$. These snapshots indicate that at high $V_{\infty}$ values, the disk exhibits very small amplitude oscillations after reaching a quasi-stable state, affecting the $QPO$ oscillation frequencies and amplitudes. This effect is evident in the left graph of Fig. \ref{QPO_diff_alpha_diff_a}, where both the frequency amplitude and the resulting $QPO$ frequencies are altered, including the disappearance of some frequency values. As previously suggested, to explain the observational results numerically, an average $V_{\infty}$ value is needed, which, according to our simulations, should be around $V_{\infty}=0.2$.

Furthermore, the right graph of Fig. \ref{QPO_diff_alpha_diff_a} shows the oscillations of the spiral waves for different spin parameters of the Kerr black hole, numerically determined when the disk is modeled with an angular velocity of $V_{\infty}=0.2$. This graph demonstrates that the black hole spin parameter does not significantly change the amplitude or frequency of the $QPOs$, with the same $QPO$ oscillation frequencies observed across these spin parameters. This consistency might result from specific factors influencing the disk behavior in this simulation, including the black hole spin, the EGB coupling constant, and the angular velocity of the perturbation.

As inferred from the right graph of Fig. \ref{QPO_diff_alpha_diff_a}, in $QPO$ production, the EGB coupling constant and angular velocity appear to dominate over the black hole spin. One reason for this could be our numerical setup, which places the inner boundary of the disk at $r=3.7 M$, whereas in Kerr black holes, the horizon is located at $<2M$.

\begin{figure*}
  \vspace{1.2cm}
  \center
  \psfig{file=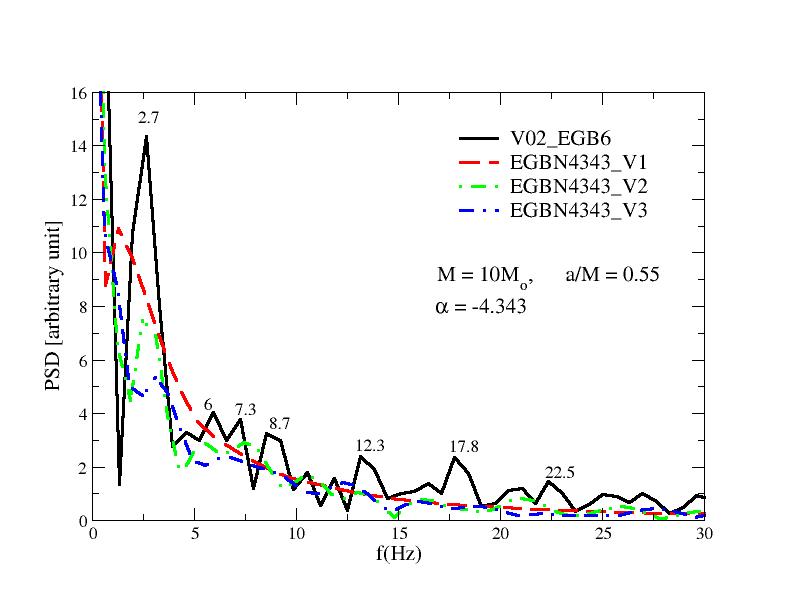,width=10cm,height=8cm} 
  \caption{Power spectrum density for models with $\alpha = -4.343$. The various lines represent the $QPO$ frequencies on the disk after perturbation with different angular velocities.
  }
\label{QPO_M3}
\end{figure*}

For large negative values of $\alpha$, the transition from initial stability to instability and then back to stability after a disruptive perturbation takes a considerable amount of time. This indicates that the model exhibits highly chaotic behavior, as discussed in previous sections and in \citet{Donmez2023arXiv231013847D}. From Fig. \ref{QPO_diff_alpha}, we see that this situation impacts the disk $QPO$ frequencies. In Fig. \ref{QPO_M3}, we have modeled the behavior of $QPO$ frequencies for large negative $\alpha$ values at asymptotic speeds of $V_{\infty}=0$, $0.2$, $0.4$, and $0.6$. The primary genuine mode always occurs, but its amplitude is much smaller for $V_{\infty}=0$, $0.4$, and $0.6$ compared to the model for $V_{\infty}=0.2$. Furthermore, for $V_{\infty}=0.4$ and $V_{\infty}=0.6$, the second genuine mode and nonlinear couplings have very small amplitudes, suggesting that the system does not achieve quasi-periodic stability necessary for generating $QPO$ frequencies. Moreover, the second mode and nonlinear couplings are completely absent at $V_{\infty}=0$, highlighting the importance of the angular velocity of the perturbing matter in the formation of low-frequency $QPOs$. These models indicate that an ideal perturbation angular velocity, particularly around $V_{\infty}=0.2$, is essential for explaining the low-frequency $QPOs$ observed, suggesting that a moderate angular velocity is required for the perturbation to align with observational data.

Finally, at the point $r=3.88M$ where PSD analyses are calculated, matter orbits the black hole at least $400$ times throughout the entire simulation. However, to establish the physical accuracy of low frequencies on a more solid foundation, simulations need to run for much longer periods. Considering the vast number of models, the complexity of the problem, and that results are already obtained with around $3\times 10^6$ time steps for each model, this seems impractical. Therefore, the low frequencies revealed in the PSD analyses in Figs. \ref{QPO_diff_alpha}, \ref{QPO_compare1}, and \ref{QPO_diff_alpha_diff_a} may need reconfirmation through long-term simulations, especially for frequencies like $2.5$, $2.6$, and $2.87$ Hz. In other words, while the low-frequency QPOs we have identified might propose a physical mechanism in the numerical simulation that could explain observational results, confirming these low-frequency QPOs may still be necessary.

\section{QPOs from Different Observed Sources}
\label{QPOs_source}

After the formation of an accretion disk around black holes, understanding the dynamic structure and oscillation characteristics of the disk is crucial. These features are key to determining the physical properties of the black hole, such as its mass and spin parameters. Observations play a crucial role in revealing the structure of the accretion disk. Information about the disk’s structure can be gleaned by examining the properties of emitted radiation across different frequency ranges within the observed electromagnetic spectrum. To characterize the oscillation properties of the disk and predict the strength of gravity, $X$-ray observations are essential. Obtaining $QPO$ frequencies from $X$-rays is a significant step in understanding the characteristics of both the black hole and the disk. However, understanding the physical mechanisms that produce the observed $QPO$ frequencies is not possible through observations alone. This is where theoretical and numerical studies become important; they can elucidate the structure of the disk, the properties of resulting shock waves, and instabilities. If numerically calculated $QPOs$ match the observed ones, it becomes possible to identify the types of instabilities, shock waves, the black hole spin parameter ($a/M$), and the EGB coupling constant ($\alpha$). In the literature, several studies elucidate the physical mechanism for observed $QPO$ frequencies \citep{Ingram2019,Kolo2020EPJC,Smith2021ApJ}. Here, we propose a different physical mechanism to explain the observed $QPO$ frequencies of various sources, thereby identifying the physical mechanism causing the $QPO$. Subsequently, we present and compare with numerical results the observed $QPO$ frequencies around black holes and black hole masses for different sources, including $X$-ray binaries and AGNs.


\subsection{QPOs from $X-$ray Binaries}
\label{QPOs_from_Xray}

$X$-ray binary systems are formed through various mechanisms in the universe. At the center of these binary star systems, there is either a black hole or a neutron star, which is surrounded by a companion star. This companion star feeds material into the black hole or neutron star, thus forming a stable accretion disk around it. In this study, we focus on $X$-ray systems with a central black hole and conduct a literature review on the $QPOs$ that result from the accretion of matter from the companion star onto the black hole, leading to the formation of a stable disk. We then discuss the results of perturbing this stable disk with certain physical parameters by modeling these systems using alternative gravity theories. We examine how the perturbed disk dynamic structure matches or differs from the observed $QPO$ frequencies from the sources discussed. We also propose possible physical mechanisms for the observed $QPOs$ in these sources.


\subsubsection{MAXI J1348-630}
\label{MAXI-J1348-630}

$MAXI$ $J1348-630$ is a low-mass $X$-ray binary located within the Milky Way galaxy. Although its physical characteristics are not yet fully understood, multiple outbursts have been observed in this source \citep{Zhang2020MNRAS}. Different types of $QPOs$ have been detected during these outbursts. Spectral analysis has led to the conclusion that there is a black hole at its center with a mass of $14.8\pm 0.9 M_{\odot}$ \citep{Titarchuk2023A&A}. The spin parameter of the central black hole has been estimated to be approximately $a/M = 0.42$  \citep{Wu2023MNRAS}. The $QPO$ frequency has been observed to vary between $0.2Hz$ and $18Hz$ as it transitions from the hard state to the soft state \citep{Zhang2020MNRAS}.

Both the calculated $a/M$ value and the observed frequency range of this source are consistent with the frequencies found in our models, which range from $2Hz$ to $25Hz$, and the black hole spin parameter used in our models. Therefore, we propose the presence of a two-armed spiral shock wave on the disk around the central black hole of this source. This shock wave might have reached a quasi-periodic phase, potentially giving rise to these $QPOs$.


\subsubsection{GX-339-4 and EXO 1846-031}
\label{GX-339-4}

$GX-339-4$ and $EXO$ $1846-031$ are black holes observed in Low-Mass X-ray Binaries (LMXBs) decades ago, with approximate masses ranging from $4M_{\odot}$ to $11M_{\odot}$. It has been observed that matter in the accretion disk around these black holes spirals towards the black hole, creating X-rays. The QPOs in these sources belong to the $Type-C$ class and occur in the low-hard and hard-intermediate states.

\citet{Zhang2023} analyzed the observational data of $GX-339-4$ and $EXO$ $1846-031$, studying the evolution of $Type-C$ $QPOs$ with spectral parameters. They found that the frequencies of these sources vary between $0.01$ and $30 Hz$ and concluded that the observed frequencies correlate with the inner radius of the disk and the mass accretion rate. This finding is in good agreement with the dynamic frequencies revealed through analytical and numerical modeling.

By analyzing the long-term $X$-ray light curves, \citet{Jin2023ApJ} studied the $X$-ray behavior over time. These analyses showed that the source transitions from the low-hard state ($LHS$) to the hard-intermediate state ($HIMS$), exhibiting $QPO$ oscillations identified as $Type-C$ during this transition. Spectral analysis further revealed that the $QPO$ frequencies range from $0.1$ to $0.64 Hz$ in the $LHS$ and from $1$ to $3 Hz$ in the $HIMS$.

Observational results for $GX-339-4$ and $EXO$ $1846-031$ reveal $QPO$ frequencies in the range of $0.1$ to $30 Hz$ in $X$-ray binaries. Earlier observations of these sources, along with these results, suggest that the $QPOs$ are likely generated by the spiraling motion of matter in the accretion disk towards the black hole \citep{Zhang2023,Jin2023ApJ}. In our numerical simulations, we perturb the stable disk around the black hole by varying specific physical parameters, such as radial and angular velocities. When the perturbation with certain angular velocities is applied to the disk around the black hole, we observe the formation of one- and two-armed spiral shock waves. Numerical results indicate that these shock waves stem from the oscillations generated by the perturbation, inducing variations in the mass accretion rate near the black hole horizon. Power spectrum analyses are conducted using mass accretion rates calculated for different EGB coupling constant ($\alpha$) values and black hole spin parameters ($a/M$).

These analyses show that the $QPOs$, especially those from perturbations with specific angular velocities as seen in Figs. \ref{QPO_diff_alpha}, \ref{QPO_diff_alpha_diff_a}, and \ref{QPO_M3}, are consistent with the observations for these sources. Conversely, as shown in the middle part of Fig. \ref{QPO_compare1}, the case where the perturbation has zero angular velocity results in the formation of shock waves in the disk, but the frequencies from the disk oscillations do not match the observations. Thus, it is determined that the perturbation with a specific angular velocity, which results in the formation of shock waves, aligns with observational data.

Consequently, the observed frequencies in these sources are closely related to the dynamical changes in the disk. The formation of shock waves on the disk, and the parameters of the perturbation, play a crucial role. This not only drives matter into the black hole but also causes the disk to oscillate, leading to strong shocks and the observed $QPOs$. The numerically observed $QPOs$ are consistent with the frequencies reported for $GX-339-4$ and $EXO$ $1846-031$ sources.


\subsubsection{GRO-J1655-40}
\label{GRO-J1655-40}

$GRO$ $J1655-40$ is an $X$-ray binary system fed by an $F$-type rotating star with a known black hole at its center. The matter ejected from the rotating star forms an accretion disk around the black hole. Along with $GRS$ $1915+105$, this source is instrumental in understanding the characteristic radio jets of AGNs. $GRO$ $J1655-40$ experiences irregular outbursts, and the mass of the central black hole is approximately $5.32 M_{\odot}$. The black hole spin has been calculated as $a/M = 0.3$, consistent with optical/NIR measurements \citep{Motta2014MNRAS}. Spectral analysis across different bands reveals $QPO$ frequencies ranging from $0.1Hz$ to $450 Hz$ in $GRO$ $J1655-40$ \citep{Remillard1999ApJ, Strohmayer2001ApJ, Belloni2012MNRAS, Motta2014MNRAS}.

The broad frequency range of $QPO$ frequencies observed in this source can be explained by the physical mechanism we propose in this article. Specifically, the two-armed shock wave, formed due to the perturbation of the disk around the black hole, traps and drives the $QPO$ modes, generating regular $QPO$ frequencies consistent with those observed in the source. Furthermore, the high frequencies that our calculations, presented in the middle part of Fig. \ref{QPO_compare1}, are in agreement with the high frequencies observed in these sources. Therefore, the physical mechanism and the angular velocity of the perturbation proposed in this article seem capable of explaining both the observed low and high frequencies.


\subsubsection{XTE J1748-288 (H1743-322)}
\label{H1743-322}

The source $XTE$ $J1748-288$ is an $X$-ray binary system that initially exhibited $X$-ray emission in a hard state before transitioning to a soft state. Observational results have shown variable $X$-ray emissions linked to a dense physical state. The gravitational pull of the black hole causes matter to infall, leading to the formation of a spiraling shock wave, which in turn, results in strong $X$-ray emissions. The frequencies of the emitted $QPOs$ vary from $0.01$ to $30Hz$ \citep{Naik2000A&A, Revnivtsev2000MNRAS} as the system transitions from the hard to the soft state. The mass of the compact object is estimated to be approximately $10$ to $15 M_{\odot}$ \citep{Revnivtsev2000MNRAS}.

The mechanism proposed to explain this source aligns completely with the mechanism we introduced in this article. The one- and two-armed spiral shock waves generated in the models, due to perturbation and their $QPOs$, are the physical mechanisms behind the observed $QPOs$ from this source. As reiterated throughout the article, it is clear that the suggested mechanism for this source requires the disk to be perturbed at a moderate angular velocity.


\subsection{QPOs from AGNs}
\label{QPOs_from_AGN}

AGNs are highly energetic events powered by the infall of matter into a massive central black hole, with masses ranging from $10^6$ to $10^9 M_{\odot}$. These events emit intense electromagnetic radiation across a wide range of time scales due to matter accretion onto the black hole \citep{Ishibashi2009A&A,Song2020A&A}. In this discussion, we explore the physical origins of the $QPOs$ occurring on the disk near the region of strong gravitational force close to the central black hole in AGNs. We examine the reasons for the emergence of different oscillation frequencies from various sources.

\subsubsection{RE J1034+396}
\label{RE_J1034+396 }

The source $RE$ $J1034+396$ was first discovered to exhibit $QPOs$ by \citet{Gierli2008Nature}. It is known for high levels of $X$-ray emission lines, observed as radiation from a nearby galaxy. Analyses of the observations and power spectrum calculations have revealed $QPO$ frequencies at the level of $2.7\times10^{-4}Hz$ from this source, marking the first detection of a strong $QPO$ frequency derived from the $X$-ray light curve for this source. The mass of $RE$ $J1034+396$ has been variously estimated; from the $H\beta$ emission line velocity distribution, it is calculated to be $6.3\times10^{5} M_{\odot}$ \citep{Kaspi2000ApJ}, while from the velocity distribution of the OIII emission, it is estimated at $3.6\times10^{7} M_{\odot}$ \citep{Tremaine2002ApJ}. \citet{Jin2020MNRAS} have listed the $QPO$ frequencies obtained from different observations of $RE$ $J1034+396$ in Table 1, varying between $2.5$ and $2.8\times10^{-4}Hz$. \citet{Czerny2016A&A} determined the mass of $RE$ $J1034+396$ using various black hole mass determination methods, suggesting a mass of $1\times10^{6} M_{\odot}$ based on integrated bolometric luminosity from a broad band of optical/UV/$X$-ray data.

Given the varied observational results indicating that the central black hole mass ranges between $6.3\times10^{5}$ and $3.6\times10^{7} M_{\odot}$, our numerical results align with the observational data. To demonstrate this agreement, calculations were performed by replacing the black hole mass of $M=10M_{\odot}$ in Figs. \ref{QPO_diff_alpha}, \ref{QPO_compare1}, \ref{QPO_diff_alpha_diff_a}, and \ref{QPO_M3} with the mass estimates for this source. Adjusting the numerical simulation frequencies for the black hole mass of $6.3\times10^{5} M_{\odot}$ shows that the $QPO$ frequencies range between $3.9\times10^{-4}Hz$ and $3.9\times10^{-3}Hz$. Consequently, it is plausible that shock waves formed in the disk of this source are responsible for the observed $QPO$ frequencies, validating the physical mechanism proposed in our article.

However, the black hole mass of $3.6\times10^{7} M_{\odot}$ cannot be reconciled with the results from our model, as it would imply $QPO$ frequencies at the micro-level. This suggests that the mass estimate derived from the $OIII$ emission line velocity dispersion may not be accurate, indicating that the central black hole of this AGN might not be as massive as initially estimated.

\subsubsection{NGC 4051}
\label{NGC_4051}

$NGC$ $4051$, located in the Ursa Major constellation, is one of the best-known galaxies and a Seyfert 1 galaxy that emits powerful $X$-rays. The observed mass of the black hole is approximately $1.7\pm0.5\times10^{6} M_{\odot}$ \citep{Denney2009ApJ}. Additionally, $QPO$ analyses conducted on this galaxy have revealed different oscillation frequencies based on various observation results, ranging between $2\times10^{-4}Hz$ and $8\times10^{-5}Hz$ \citep{Vaughan2011MNRAS, McHardy2006Nature}.

If the black hole mass, $1.7\pm0.5\times10^{6} M_{\odot}$, is used in Figs. \ref{QPO_diff_alpha}, \ref{QPO_diff_alpha_diff_a}, and \ref{QPO_M3} instead of $M=10M_{\odot}$, the frequencies are observed to be in good agreement with both numerical and observational $QPOs$. In the numerical simulation, the frequency for this black hole mass ranges between $1.4\times10^{-5}Hz$ and $1.4\times10^{-4}Hz$. The $QPO$ frequencies obtained from numerical modeling align with the observed frequencies of this source, strongly suggesting that the accretion disk around this black hole hosts spiral shock waves, which could be the cause of the observed $QPOs$.

\subsubsection{1H 0707-495}
\label{1H 0707-495}

$1H$ $0707-495$ is a source that exhibits $X$-ray oscillations on very short time scales. Therefore, it is believed that the emitted $X$-rays originate in a region very close to the black hole. Consequently, the physical parameters of the black hole, such as its spin parameter, mass, and EGB coupling constant, along with the physical parameters of perturbations falling toward the black hole, are thought to influence the amplitude and frequency of the oscillations. Due to strong gravitational attraction, the $X$-ray spectrum exhibits more chaotic behavior. This source is classified as a Narrow-Line Seyfert 1 galaxy, characterized by a very narrow emission line spectrum. Analysis of observational data has revealed that the mass of the central supermassive black hole is on the order of $2\times10^{6} M_{\odot}$ or $4\times10^{6} M_{\odot}$ \citep{Done2016MNRAS}. Following the $X$-ray spectrum analysis of $1H$ $0707-495$, in addition to the previously found $QPO$ frequency of $\sim 1.2\times 10^{-4}Hz$, a new $QPO$ frequency of $\sim 2.6\times 10^{-4}Hz$ has been discovered \citep{Zhang2018ApJ}.

The observed frequency of this source is believed to be produced very close to the black hole, where gravity is exceptionally strong. Since the numerical $QPOs$ are calculated using data obtained at $r = 3.88 M$, the properties of this source can be elucidated. The $QPOs$ obtained in the models are in good agreement with the observed $QPOs$ of this source. The frequency range obtained in the numerical simulation for the  black hole with a mass of $2\times10^{6} M_{\odot}$ varies from $\sim 1\times 10^{-5}Hz$ to $\sim 1.25\times 10^{-4}Hz$. Therefore, we suggest the presence of a two-armed spiral shock wave around the central massive black hole in this source, which has generated these frequencies through regular $QPOs$.


\section{Discussion and Conclusion}
\label{Conclusion}

Perturbing stable accretion disks around black holes and calculating $QPOs$ after the formation of $QPOs$ can help explain the source of $QPOs$ observed in $X$-ray binaries and AGNs. Uncovering the nature of shock waves on the accretion disk, generated through numerical modeling, is crucial. This involves understanding the dynamic structure of the shock wave, its dependence on parameters such as $a/M$ and $\alpha$—which play a crucial role in modified gravity theories that include higher-order curvature terms—and identifying the physical parameters of the perturbing matter. By doing so, we grasp the nature of the physical factors that lead to the shock wave on the disk and its $QPOs$, thereby providing a mechanism to explain the observed $QPO$ frequencies in the aforementioned systems.

The effect of $\alpha$ on the energy flux, radiation spectrum, and last-stable orbit in thin accretion disks is theoretically explained in \citet{HeydariFard2021EPJC}. In our study, we conduct numerical modeling with different values of $a/M$, $\alpha$, and the angular velocity of the perturbation, denoted as $V_{\infty}$. Our aim is to reveal the oscillation characteristics and nature of the shock wave under these varying parameter settings. We numerically calculate the behavior of the disk immediately after perturbation, the mass loss rate, and the transition to $QPOs$ after the shock wave forms. We also examine the instability and its intensity on the disk. Subsequently, we discuss the stability transitions of the $m=1$ and $m=2$ modes by calculating their mode powers. These modes in the numerical analysis provide insights into when the disk reaches saturation, whether it exhibits $QPOs$ after saturation, and how the modes vary according to different physical parameters \citep{Montero2004MNRAS}.

The numerical results show that the $EGB$ coupling constant, $\alpha$, which defines the spacetime metric, affects the dynamics and oscillations of the shock wave formed on the disk. For most $\alpha$ values used in the models, the shock wave response to $\alpha$ remains fairly consistent. However, significant changes in the disk structure and oscillation behavior are observed for large negative values of $\alpha$. Conversely, it has been concluded that the black hole spin influence on shock waves is not as pronounced as that of $\alpha$. This is likely because the numerically modeled accretion disk has an inner radius at $r=3.7M$, while for spinning black holes, the horizon is located at $r<2M$. This suggests that the effect of the spin parameter, which is more pronounced near the black hole horizon, diminishes with increasing distance.

The analysis of results from numerical models shows that the angular velocity of the perturbation, which disturbs the disk near the black hole, significantly affects various aspects. These include the disk dynamic structure, the accreted matter around the black hole, the emergence of Papaloizou-Pringle instabilities, the quasi-periodic behavior of the disk, and consequently, the numerically observed $QPO$ frequencies. The parameter defining the perturbation angular velocity is denoted as $V_{\infty}$. In cases where $V_{\infty}$ is small (including zero) or very large, shock waves still form on the disk, but its dynamic structure and oscillation properties change. For large $V_{\infty}$, the shock wave appears more compact and exhibits a low oscillation amplitude. Conversely, for small $V_{\infty}$, the shock wave oscillation is larger, and it does not exhibit fully quasi-periodic behavior.

However, for $V_{\infty} = 0.2$, the shock wave formed on the disk exhibits quasi-periodic behavior. Power spectrum analyses indicate that at an angular velocity proportional to $V_{\infty} = 0.2$, $QPO$ frequencies are observed, aligning with the $X$-ray binaries and AGNs discussed in the article. Therefore, the disk structure and the two-armed shock wave generated at this $V_{\infty} = 0.2$ are proposed as physical mechanisms to explain the observed $QPOs$ in these sources. Since the numerically calculated $QPOs$ are independent of the black hole mass, the numerical results explain the low-frequency oscillations in $X$-ray binaries and AGNs.

Finally, our numerical findings indicate that $QPOs$ are more influenced by the $EGB$ constant than by the black hole spin parameter ($a/M$). However, we emphasize that the primary factor affecting oscillations and $QPOs$ is the perturbation angular velocity. Our models show that the perturbation asymptotic velocity at $V_{\infty} = 0.2$ generates $QPO$ frequencies independently of the black hole spin and the $EGB$ coupling constant. Thus, we propose that, for a moderate value of $V_{\infty}$, the formation of a two-armed spiral shock wave on the disk serves as a crucial mechanism for explaining low-frequency $QPOs$.



\section*{Acknowledgments}
I would like to express my sincere gratitude to the anonymous referee for their meticulous review and
insightful feedback. Their constructive comments significantly improved the quality of this paper.
All simulations were performed using the Phoenix  High
Performance Computing facility at the American University of the Middle East
(AUM), Kuwait.\\

\bibliography{paper.bib}

\end{document}